\documentclass[12pt]{article}

\newenvironment{frontmatter}{}{}
\renewenvironment{abstract}{\vspace{-0.8cm}\begin{center}\begin{minipage}[h]{0.9\textwidth}\centerline{\bf Abstract}\vspace{2mm}\par }{\end{minipage}\end{center}}
\newenvironment{keyword}{\begin{center}\begin{minipage}[h]{0.9\textwidth}\small{\em Key words:\/}}{\end{minipage}\end{center}\vspace{0.5cm}}
\newenvironment{ack}{\vspace{0.7cm}{\bf Acknowledgments}\par\vspace{0.3cm}}{\vspace{0.3cm}}
\newcommand{\PACS}[1]{\par{\em PACS:\/} {#1}}
\newcommand{\figheight}{7.2cm}
\newcommand{\BibTitle}[1]{}

\newcommand{\refeq}[1]{(\ref{#1})}  
\newcommand{\mcite}[2]{\cite{#2}}   
\newcommand{\up}[2]{$^{#1#2}$}    

\newcommand{\Wav}[1]{\langle W_{#1} \rangle} 
\newcommand{\EpJT}[1]{\Delta E_{#1}}  
\newcommand{\Imax}{I_\mathrm{m}}    
\newcommand{\Qspec}{Q_\mathrm{spec}} 
\newcommand{\Ttot}{T_\mathrm{tot}}  
\newcommand{\Pbar}[1]{${\bar P}_{#1}$}      
\newcommand{\spt}{K^\mathrm{d}}          
\newcommand{\sptJT}[1]{K_{#1}^{\mathrm{d}}} 

\input{psfig}
\begin{document}

\begin{frontmatter} 
\title{Pairing Interaction in N=Z Nuclei\\ with Half-filled High-$j$ Shell}


\author{A.Juodagalvis\thanks{e-mail: andriusj@matfys.lth.se} \\ 
		\footnotesize
  Mathematical Physics Division, Lund Institute of Technology,\vspace{-2mm}\\
		\footnotesize
  S-22100 Lund,  Sweden}
\date{}
\maketitle

\begin{abstract}
The role of $L$=0 pairing interactions (both $T$=0 and $T$=1) in three
selected bands with the total isospin $\Ttot$=0 in \up22Na,
\up48Cr and \up90Rh nuclei is discussed in the spherical shell
model. These bands were selected requiring termination in a most
unfavourable way.  The investigated effects coming from pairing
interactions include the change of nuclear wavefunction, quadrupole
properties and state energy. To follow the gradual change of nuclear
properties, the investigation is performed in a perturbative approach
by defining the Hamiltonian with a variable content of the pairing
interaction. It is shown that the pairing interaction does not affect
the quadrupole properties, if the energy distance to other close lying
states is large. The calculated pairing energy is shown to have a
similar spin dependence in these three bands. In addition, the
influence of the model space as well as of the residual interaction on
the derived pairing energy is discussed.
\end{abstract}

\end{frontmatter}

\section{Introduction}

The pairing force plays an important role for nuclear properties like
binding energy, deformation, moment of inertia. It is a long standing
question what is the quantitative role of the isoscalar ($T$=0) and
isovector ($T$=1) pairing interactions. Their competition may be
observed in the double-odd N=Z nuclei, where the lowest states with
the total isospin $\Ttot$=0 and 1 have very similar binding
energy, and the transition from the $\Ttot$=0 to
$\Ttot$=1 ground state, as the mass number increases, is seen.
This phenomenon has recently regained attention, see e.g.\
\cite{MartinezLangankeNPPairing} and references herein.

To investigate the role of the pairing interactions, many attempts
were made to incorporate those correlations in the
single-particle models like 
Hartree-Fock (HF) approximation (see e.g.\
\cite{Goodman,Civitarese,Wyss,Goodman99}). In the HF approach, the pairing
correlations should be added. A different situation occurs in the
shell model which takes into account the residual interaction
completely.  Here the pairing force is a part of the full interaction,
and in principle it cannot be separated from the latter. It is known
that the shell model Hamiltonian, consisting of the single particle
term and the $T$=1 pairing and quadrupole interactions
\cite{PairingPlusQuadrupoleModel}, gives a reasonable description of
nuclear properties. However, a question whether this is a sufficient
minimal shell model Hamiltonian appeared again, when the theoretical
background for strengths of pairing and quadrupole forces was
reconsidered and it was shown that a schematic $T$=0 pairing is also
an important part of the realistic interaction \cite{DufourZuker}.
The interest was also stimulated by the availability of advanced shell
model calculations (see e.g.\
\cite{MartinezLangankeNPPairing,PovesMartinezCr48}) as well as by
new experimental data available (e.g.\ \cite{LenziTalk,Ga62Exp}).

This paper discusses the role of a schematic $L$=0 pairing. The
two-particle states with spin $S$ and isospin $T$ equal to $ST$=01 or 10
are the ones favoured by Wigner's SU(4)
classification. Those states are also among the few lowest eigenvalues
of the multipole part of a realistic Hamiltonian
\cite{DufourZuker}. The real nuclear interaction breaks SU(4)
symmetry (e.g.\ by spin-dependence). However, it is not completely
destroyed, and its presence shows up most clearly in the lighter nuclei
through selection rules for $\beta$-decay. 

The influence of the isoscalar and isovector $L$=0 pairing components
of the effective interaction has been discussed for the nuclei in the
middle of $pf$-shell
\cite{MartinezLangankeNPPairing,PovesMartinezCr48,AJSAGa62}. The present paper
refers to three N=Z nuclei having half-filled high-$j$ shell. These
three selected nuclei (\up22Na, \up48Cr and \up90Rh) are of particular
interest. First of all, since protons and neutrons are occupying same
shells, the neutron-proton interaction plays an important
role. Another interesting feature is a rotational behaviour of the
bands that terminate at spin $\Imax$=($p$+1)$^2\hbar$ ($p$ is the
principle quantum number of the harmonic oscillator shell). Those
bands show backbending in the spin region ($\Imax-$4,$\Imax-$2). They
terminate in a spherical state, since all components of the high-$j$
shell orbits having positive angular momentum projection are occupied.
In addition, they terminate in a ``most unfavoured'' way
\cite{AJSABandTermination}, since last two transitions before the
termination involve breaking and alignment of two $T$=1 pairs (one
neutron pair and one proton pair). Besides these three nuclei, such a
band termination may occur in other nuclei (for example, an excited
positive parity band in \up47Cr, having one hole in the $sd$-shell),
but those bands would lie at higher excitation energy.

The most unfavoured band termination was the criterion to select
particular bands in these three nuclei. These bands
have the total isospin $\Ttot$=0. The role of pairing
interactions was investigated using the shell model. For two nuclei,
\up22Na and \up48Cr, experimental data exists. But no data is
available for \up90Rh, thus we have to rely on the ability of the
chosen interactions to describe properties of the neighbouring
nuclei. The aim of this paper is to show that the selected bands have
some features in common, and that the $L$=0 pairing interactions
change the properties of the bands in a similar way. 
The pairing influence to the nuclear wavefunction and
quadrupole properties is also discussed.

The paper is organized in the following way. Computational details and
comparison with experimental data are both discussed in
section~\ref{sect-ApproachDescription}. Similarities in the selected
bands as well as the predicted properties of the selected band in
\up90Rh are also discussed in a greater detail there. At the end of
the section, evaluation procedure of the role of pairing interaction
is explained. The results are discussed in section~\ref{sect-Results}.
It is shown that the $L$=0 pairing interaction affects the behaviour
of the selected bands in a similar way. And finally, conclusions are
drawn in section~\ref{sect-Summary}.

\section{Description of the approach}
\label{sect-ApproachDescription}

This section discusses three selected bands in a greater detail
(subsection \ref{ssect-Selection}). Then computational details and
comparison with experimental data are given (subsection
\ref{ssect-Experiment}). Finally, an approach to extract the role of
pairing interaction is explained (subsection \ref{ssect-Approach}). The
obtained results are discussed later in section \ref{sect-Results}.

\subsection{Bands having most unfavoured termination}
\label{ssect-Selection}

Among the N=Z nuclei, there is a special class of nuclei having
exactly half-filled high-$j$ shell. A peculiarity in these nuclei is
that it costs more energy to construct the state, where all the
valence particles are aligned in this high-$j$ shell giving maximum
spin $\Imax$=($p$+1)$^2$ (where $p$ is the principle quantum number),
than a rotational trend would suggest~\cite{AJSABandTermination}.
This effect comes mostly from $JT$=01 pairing interaction. It is
interesting to see that there are similarities among the bands also in
the spin dependence of the content of pairing energy.

\begin{figure}[tb]
\centerline{\psfig{figure=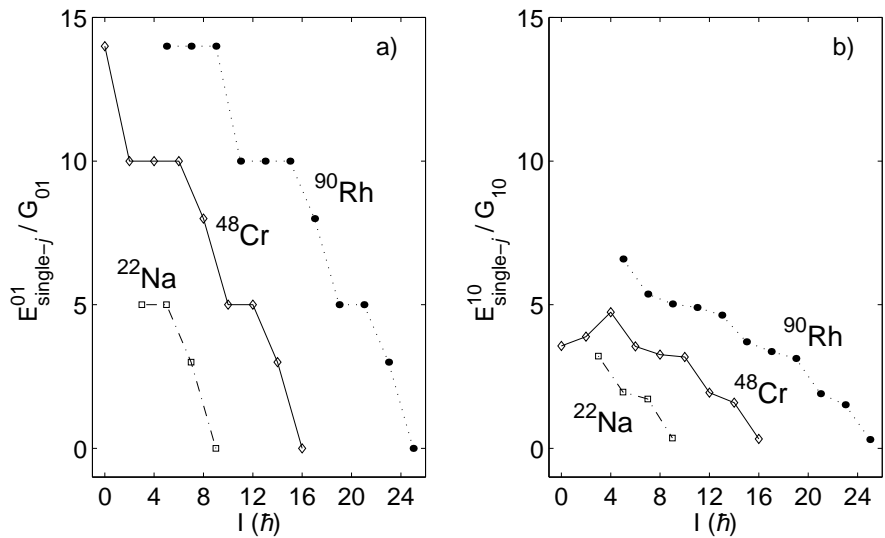,height=\figheight}}
\caption{
Spin dependence of the $JT$ pairing energy in the lowest
$\Ttot$=0 states in the model space consisting of a high-$j$
shell only. The $JT$=01 case is shown on the left, while $JT$=10 case
is shown on the right. Formula for the $JT$=01 pairing energy,
parameterized in terms of the total seniority and reduced isospin, is
taken from \cite{deShalit}. }
\label{fig-SingleJEp}
\end{figure}

In this paper, three N=Z nuclei are discussed, namely, \up22Na, having
6 valence particles (3 protons and 3 neutrons) in the $d_{5/2},$
\up48Cr, having 8 particles in $f_{7/2},$ and \up90Rh, having 10
particles in $g_{9/2}$ shell, when the single-particle level ordering
from a harmonic oscillator potential with $\ell^2$ and $\ell\cdot s$
terms is assumed. In the single-$j$ shell model with the $JT$=01
pairing interaction only, bands leading to the maximum spin state
behave in a similar way (see fig.\ref{fig-SingleJEp}$a$). Let $\Imax$
is the spin of a completely aligned state (which should be spherical
since all orbits with positive $\vec j$ projections are occupied):
$\Imax$=9$\hbar$ for \up22Na, 16$\hbar$ for \up48Cr, and 25$\hbar$ for
\up90Rh. Let us discuss only the lowest states at spins $\Imax-$2$k$
where $k$ is an integer number. These states have the total isospin
$\Ttot$=0. The state with spin $\Imax-$2 would be a linear
combination of states where one pair of identical particles is coupled
to angular momentum 0, and the spin is obtained from the remaining
unpaired particles. Similarly, the $\Imax-$4 state would have two
pairs of particles (one neutron pair and one proton pair) coupled to
angular momentum 0. For \up22Na, that is the maximum spin state
produced by two odd particles (5$^+$, bandhead $K^\pi$=3$^+$). Since there
is no change in the pairing energy when going from $\Imax-$4 to
$\Imax-$6 state, the $\Imax-$4 state may be somewhat irregular in the
real energy spectrum. In the discussed cases there is a backbending in
\up48Cr and \up90Rh, but in general the change of the moment of
inertia may be less pronounced. The state with spin $\Imax-$10 (having
10 units of $JT$=01 pairing energy in a single-$j$ model, see
fig.\ref{fig-SingleJEp}$a$) would be a linear combination of states
where 6 particles contribute angular momentum zero: two pairs of one
kind and one pair of the other kind of particles are all contributing
angular momentum 0. Since at least 6 valence particles are not
contributing to spin, the irregularity in the energy spectrum may be
not seen unless this $j$-shell is dominating. The state $\Imax-$16
would have 4 angular-momentum-zero pairs. This state is the ground
state in \up48Cr, while in \up90Rh it is the maximum spin obtained
from odd proton and neutron (9$^+$, bandhead $K^\pi$=5$^+$).

There are more bands terminating in a most unfavourable way. One can
think of lighter nuclei with half-filled high-$j$ shell of harmonic
oscillator major shells having principle quantum number $p$=1 or 0
($^{8}$Be and $^{2}$H, respectively), but there are very little states
before the maximum spin state, thus no bands are formed. A more
realistic examples are excited bands in nuclei close the three earlier
mentioned nuclei. First example could be one-particle-one-hole (1p-1h)
bands of unnatural parity in nuclei having mass number
($A_\mathrm{m}-$1) (here $A_\mathrm{m}$ is the mass number of any of
the three middle high-$j$ shell nuclei). For example, an excited band
in \up21Ne (or \up21Na), where one proton (neutron) is excited from
the $p$- to $sd$-shell. This band would terminate at spin $I$=9+$j_h$,
where $j_h$ equals 1/2 or 3/2 depending on the position of the hole in
$p$-shell. Other examples are \up47Cr (or \up47V) and \up89Ru (the
band in \up89Rh is less likely to be observed, at least in the nearest
future). One can also think of bands based on 2p-2h excitation in
($A_\mathrm{m}-$2) mass nuclei, and so on.

The most unfavoured band termination may also be encountered in cases
when the last transition before the band termination involves
alignment of a pair (coupled to angular-momentum-zero) that
brings only 2$\hbar,$ and the relative difference of neutron and proton
seniorities (i.e.\ the reduced isospin) decreases. In the cases discussed 
in this paper, the reduced isospin is zero at spin $\Imax-$4, then it
increases at state $\Imax-$2 and decreases again at the aligned
state with the total angular momentum $\Imax$.

The discussion of pairing features in a single-$j$ shell is based on
the fact that the high-$j$ shell dominates in the cases discussed
here, i.e.\ either it has the largest occupancy and/or its change has
greater impact to the band properties. For example, shell occupancies
in some states of \up48Cr are shown in fig.\ref{fig-Cr48I6Occupancies}
(occupancies in all yrast states of \up48Cr may be found in
ref.\cite{SMCHFBCr48}). The shell occupancies along the selected band
in \up90Rh are shown in fig.\ref{fig-Rh90Occupancies}.

The role of $JT$=01 pairing interaction is well acknowledged. The
seniority quantum number led to an additional classification of
many-particle configurations. While the role of $JT$=10 pairing
interaction is less pronounced: models without this interaction
successfully describe nuclear properties, and there is no additional
(and useful) classification in terms of $JT$=10 coupled pairs.
Although in the realistic interactions $JT$=10 pairing is stronger
than $JT$=01 \cite{DufourZuker}: the ratio of the strength parameters
$G_{10}/G_{01}$$\approx$1.5, see e.g.\ Table~\ref{tbl-GValues}; the
features of this pairing interaction in a single-$j$ shell (see
fig.\ref{fig-SingleJEp}$b$) are less prominent in the calculated or
observed spectrum. For example, the behaviour of $JT$=10 energy in a
single-$j$ model predicts an increased energy loss between states
having spins $\Imax-$6 and $\Imax-$4. In the observed spectra
(fig.\ref{fig-NaCrEnergies}) as well as in the full calculations
(figs.\ref{fig-NaCrEnergies},\ref{fig-RhEnergies}), there is a
backbending between those states, i.e.\ the behaviour is opposite to
the predicted from the $JT$=10 pairing energy spin-dependence in a
single high-$j$ shell. It may be noted, however, that the unequal
energy spacing between states with spins $\Imax-$4, $\Imax-$2 and
$\Imax$ is predicted by $JT$=10 as well as by $JT$=01 pairing
interaction.

One reason for the small relative importance of $JT$=10 pairing may be
the fact, that the interaction matrix elements of $JT$=10 pairing are
smaller than those of the $JT$=01: for example, in the single-$j$
case, the ratio of the matrix elements is $G_{10}/G_{01}\, (j+1)/(3j)$
which is less than 1 for $j$$>$$\frac12$. The pairing energy would be
larger if the high-$j$ shell spin-orbit partner would be included in
the model space, as discussed below in subsection
\ref{ssect-ModelSpace}. However, that does not change much the spin
dependence of the pairing energy (see fig.\ref{fig-Epairing}). Thus
the main reason, why the $JT$=10 pairing interaction was not found
being important, may be its smooth spin dependence (see
fig.\ref{fig-SingleJEp}$b$). Small irregularities suggested by
fig.\ref{fig-SingleJEp}$b$ are likely to be found only in the cases
when $JT$=01 pairing energy does not change. Another possibility could
be the cases when $JT$=01 pairing is blocked, e.g.\ in the doubly-odd
nuclei \cite{AJSAGa62}. In subsection
\ref{ssect-ResidualInteraction} it is discussed that the $JT$=10
pairing interaction may be responsible for the favouring of $I$=7 state
in \up90Rh calculated using SSSV interaction.

The discussion above was referring to the pairing energy spin
dependence along {\em one\/} band, and is continued in subsection
\ref{ssect-PairingEnergy}. One may find a more general discussion about
the relative importance of $JT$=01 and $JT$=10 pairing energies in
the shell model framework in ref.\ \cite{MartinezLangankeNPPairing}.

\subsection{Computational details and comparison with experiment}
\label{ssect-Experiment}

In present work, shell model calculations were performed using the
computer code ANTOINE~\cite{Antoine}. The residual interactions,
used in the calculation, were:
USD~\cite{USD} interaction in the full $sd$-shell for \up22Na,
KB3~\cite{KB3} in the full $pf$-shell for \up48Cr, Gross and Frenkel
interaction \cite{Frenkel} (referred to as GF) as well as SLGT0
\cite{SLGT0} in $p_{1/2}g_{9/2}$, and Sinatkas et al.\ \cite{Sinatkas}
(referred to as SSSV) in $g_{9/2}p_{1/2}p_{3/2}$
for \up90Rh. Electromagnetic quadrupole properties were
calculated using the effective charges $q_\pi$=1.5$e$ for protons and
$q_\nu$=0.5$e$ for neutrons.

\begin{figure}[tb] %
\centerline{\psfig{figure=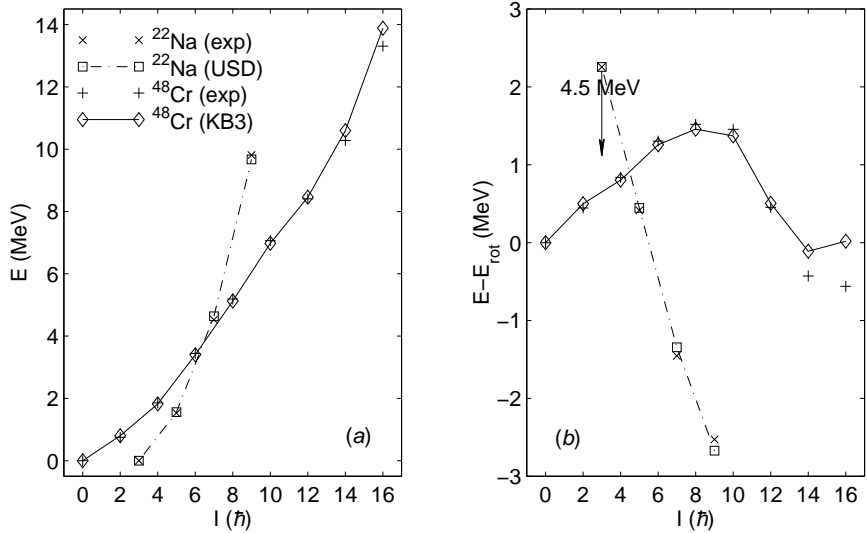,height=\figheight}} %
\caption{%
Energies of selected bands in \up22Na and \up48Cr. For the calculated
states, the residual interaction is given in parentheses. Calculated
spectra of \up22Na and \up48Cr were reported in refs.\ \cite{USD} and
\cite{A48SMcalc}, respectively. Plot $a$ shows excitation
energies, and plot $b$ shows the same energies but with a rotational
reference, $E_\mathrm{rot}$=32.32$A^{-5/3}I$($I$+1), subtracted. In
the latter plot, 4.5 MeV was added to all energies in
\up22Na to make the figure more compact, that is also indicated by an
arrow.}
\label{fig-NaCrEnergies} 
\end{figure}

\begin{figure}[tb] %
\centerline{\psfig{figure=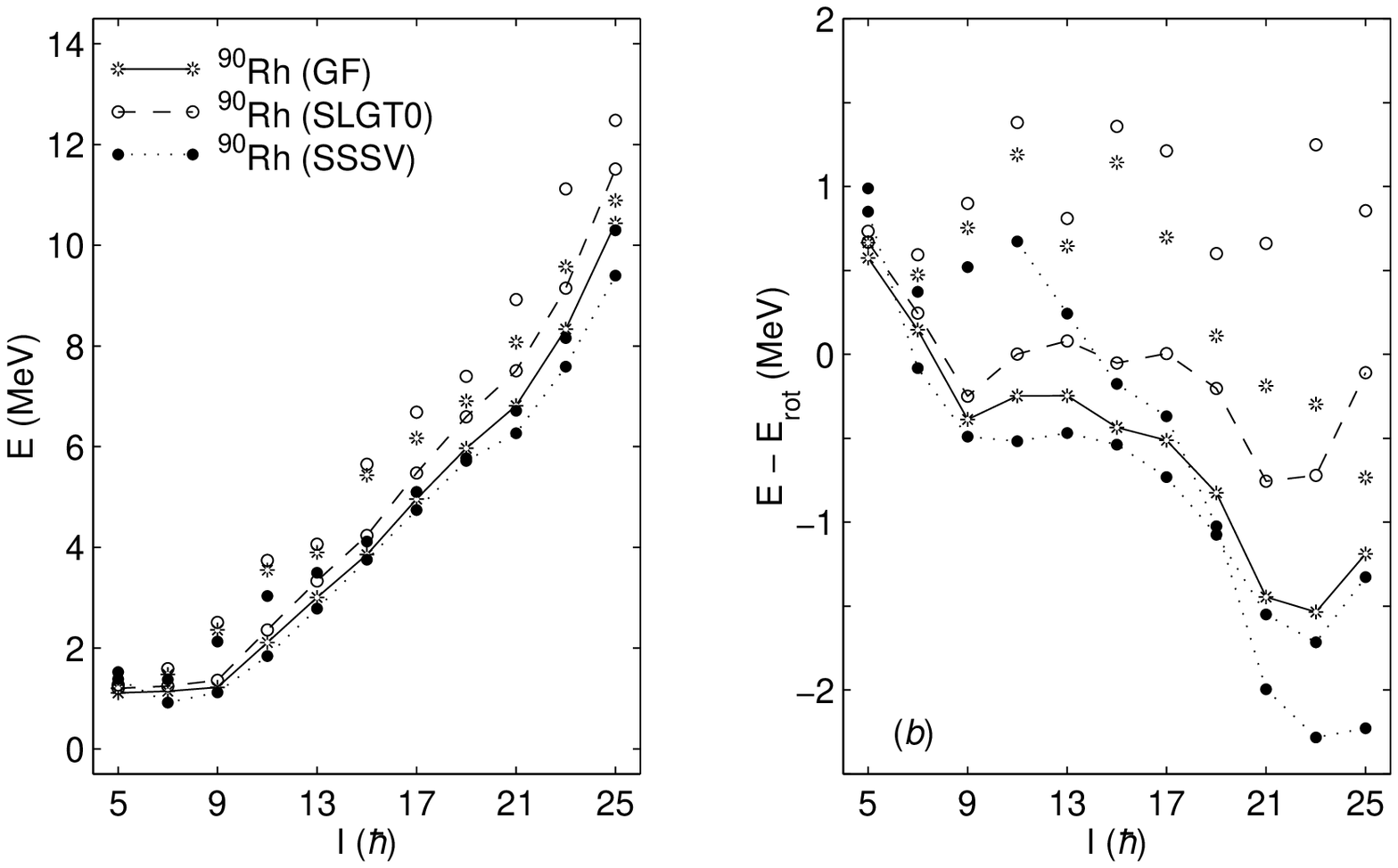,height=\figheight}} %
\caption{%
Energies of odd-$I$ members of $K^\pi$=5$^+$, $\Ttot$=0 band in
\up90Rh calculated with GF, SLGT0 (cf.\cite{SLGT0}) and SSSV
interactions.  For explanations see fig.\ref{fig-NaCrEnergies}. Yrast
states are connected by lines, while second excited states for each
spin value are indicated by a corresponding symbol not connected by a
line, except for the SSSV results shown in plot $b$, where the states
forming bands (having large $E2$ transition strengths) are connected
by dotted lines.}
\label{fig-RhEnergies} 
\end{figure}

Fig.\ \ref{fig-NaCrEnergies} presents a comparison of the calculated
and experimental energies of the selected $\Ttot$=0 states in \up22Na
and \up48Cr: These are odd-spin members of the ground state band
having $K^\pi$=3$^+$ in \up22Na (experimental data taken from
\cite{NaExpData}), and the ground state band having $K^\pi$=0$^+$ in
\up48Cr (experimental data taken from \cite{CrExpData}). As can be
seen, the agreement between experimental data and calculated values is
excellent. There is no experimental data available on \up90Rh. In
fig.\ref{fig-RhEnergies} the calculated odd-spin members of the
$K^\pi$=5$^+$ band in \up90Rh are shown. Those two figures (1 and 2)
show, that after a rotational reference $E_\mathrm{rot}$ is
subtracted, the bands tend to bend up close to the band termination
(so-called ``unfavoured band termination'' \cite{Afanasjev}). In
addition, the state $\Imax-$4 deviates from a regular behaviour of the
preceding states (not seen in \up22Na, since there is only one such
state). The irregularity at spin $\Imax-$10, discussed in subsection
\ref{ssect-Selection} as being possible to occur, is not seen in
\up48Cr and \up90Rh at $I$=6 and 15, respectively. The exceptions
are GF and SLGT0 calculations of \up90Rh, where a very restrictive
model space allows rather poor shell mixing. The calculated
probability of a configuration, where all active particles are in the
high-$j$ shell producing spin $\Imax$, is large: $>$90\% for \up48Cr
as well as
\up90Rh (SLGT0 and SSSV calculation), 54\% for \up22Na, and 62\% for
\up90Rh (GF).

\begin{figure}[tb]
\centerline{\psfig{figure=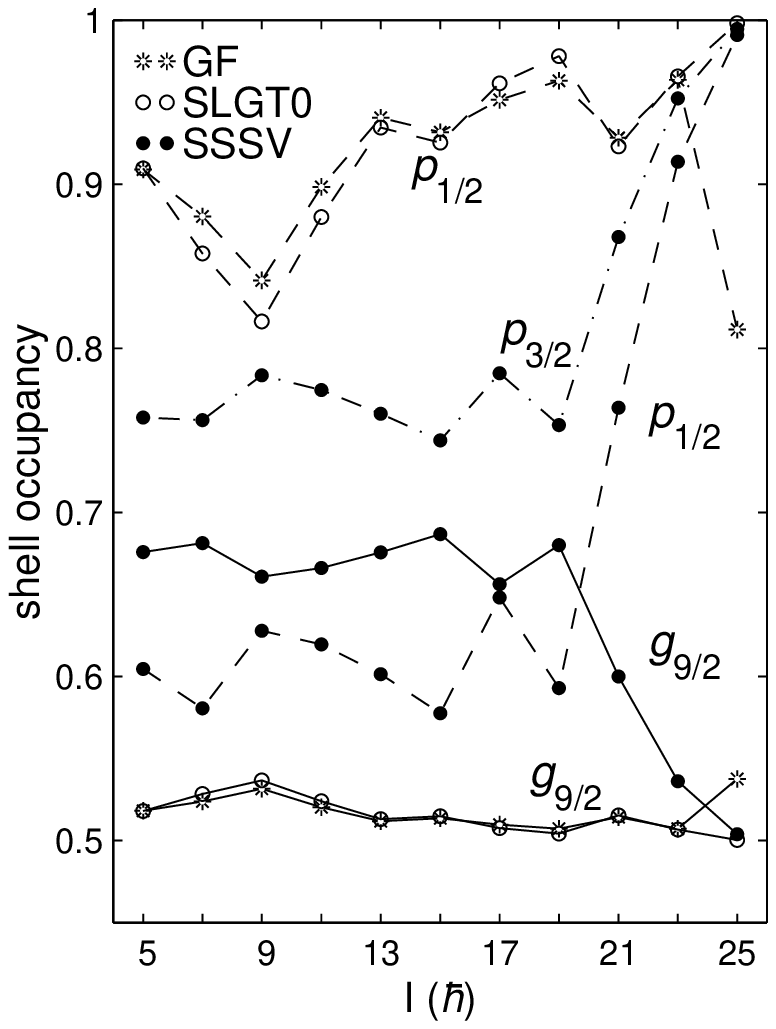,height=\figheight}}
\caption{Calculated shell occupancies in $K$=5$^+$ band in \up90Rh. 
Three residual interactions were used: GF (marked by stars) as well as
SLGT0 (open circles) in the configuration space
$(p_{1/2}g_{9/2})^{14}$, and SSSV (filled circles) in
$(g_{9/2}p_{1/2}p_{3/2})^{-10}$. The $j$-shell is indicated close to a
corresponding line. In the SSSV case, the number of holes was
converted to the number of particles in shells.  }
\label{fig-Rh90Occupancies}
\end{figure}

\begin{figure}[tb]
\centerline{\psfig{figure=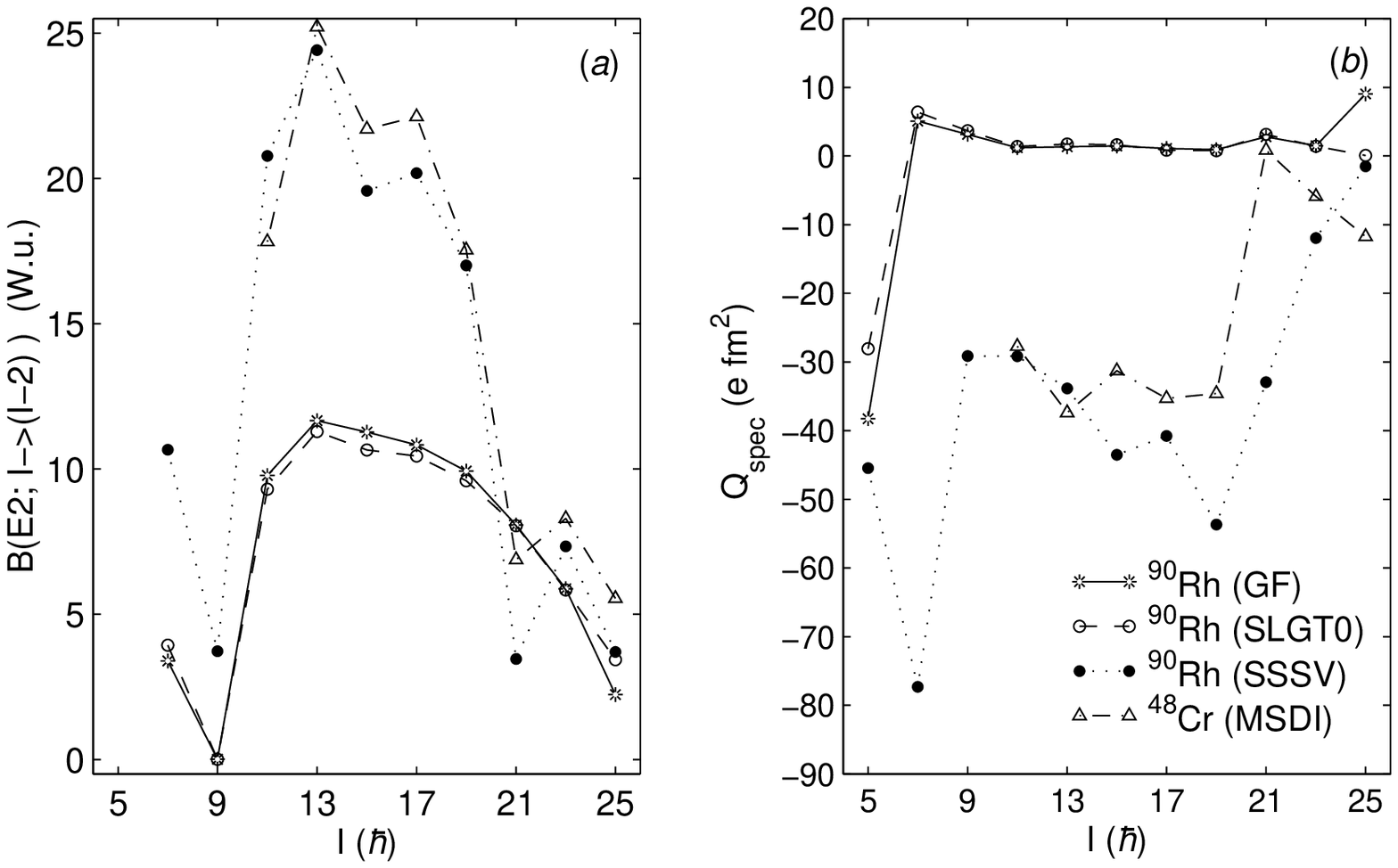,height=\figheight}}
\caption{%
Calculated $E2$ properties along the $K^\pi$=5$^+$ band in \up90Rh and
the ground state band in \up48Cr: strengths of stretched $E2$
transitions, $B(E2),$ are shown on the left; and spectroscopic
moments, $\Qspec\,,$ are shown on the right. Values for \up90Rh were
calculated using the GF (marked by stars), SLGT0 (open circles) and
SSSV (filled circles) interactions. The modified surface-delta
interaction (referred to as MSDI and marked by triangles) in the model
space $(f_{7/2}p_{3/2})$~\cite{AJSABandTermination} was used for
\up48Cr, and the spin values for \up48Cr are shifted by 9$\hbar$. }
\label{fig-Rh90E2}
\end{figure}

Since there is no experimental data on \up90Rh available, only
calculated states are shown in fig.\ref{fig-RhEnergies}. Calculations
were performed in the configuration spaces ($p_{1/2}g_{9/2}$)$^{14}$ with GF
and SLGT0 (cf.\cite{SLGT0}) interactions, and
($g_{9/2}p_{1/2}p_{3/2}$)$^{-10}$ using SSSV interaction. These three
residual interactions predict that the ground state has spin $I$=0 and
the total isospin $\Ttot$=1, in agreement with the mass-dependence of
the both spin and isospin in the ground state of doubly-odd nuclei
\cite{OddOddGroundState}. The excitation energy of the
lowest $\Ttot$=0 states is predicted to be $\sim$1.3~MeV.  The SSSV
interaction predicts almost spherical ground state of \up90Rh, with
the deformation $\beta$$\approx$0.13 (value estimated from $B(E2;
2_1^+\rightarrow 0_1^+, \Ttot=1)$, assuming axially symmetric rotor
model is valid). The interactions, however, predict slightly different
behaviour of the band of interest. For example, the SSSV interaction
favours $I^\pi$=7$^+$ state as compared to the 5$^+$ state. In
addition, energies calculated in a more restricted model space (with
the GF and SLGT0 interactions) are more irregular than those 
obtained using the SSSV interaction, see fig.\ref{fig-RhEnergies}$b$. Like in
\up48Cr case at spin $\Imax-$10=6$^+$, there is no irregularity at
spin $\Imax-$10=15$^+$ in \up90Rh according to the SSSV
calculation. That may be explained by the shell mixing (see occupation
numbers in fig.\ref{fig-Rh90Occupancies}). The band terminates at spin
$\Imax$=25$^+$ in an almost spherical shape, with a half-filled
$g_{9/2}$ shell. An exception is the GF calculation where the
terminating state contains 62\% probability of configuration
$(p_{1/2}^2g_{9/2}^5)^2$ and 37\% of $(p_{1/2}^1g_{9/2}^6)^2,$ and the
nucleus has an oblate shape. The band termination is also indicated by
a decrease in $B(E2)$ values (see fig.\ref{fig-Rh90E2}$a$). All three
interactions suggest that the $K^\pi$=5$^+$ band in \up90Rh backbends
at spin $\Imax-$6=19$^+$.  Backbending will be discussed after
comparison of the calculated occupation numbers and $E2$ properties.

The calculated $j$-shell occupancies (average number divided by the
maximum number of particles in the shell) in \up90Rh are shown in
fig.\ref{fig-Rh90Occupancies}. To get particle occupancies in the
SSSV calculation, the average number of holes was converted to the
average number of particles. As can be seen from the figure, the
$p_{1/2}$ shell is filled almost all the time (occupancy is greater
than 90\% for most spin values) in the GF as well as SLGT0
calculations, resulting in a half-filled $g_{9/2}$ shell for all band
states (except for the spin $\Imax$ state in the GF calculation, as
mentioned above). While in the SSSV calculation, this shell contains
more than 10 particles (is more than half-filled) for spins
$I$$<$21. The mixing reduces for $I$$\geq$21, when 6 particles
aligned in the $g_{9/2}$ shell may contribute 21$\hbar$.
Note that an increase of the $f_{7/2}$ shell occupation by
particles, approaching the aligned state $\Imax$=16 in \up48Cr
\cite{SMCHFBCr48}, corresponds to an increase of the $g_{9/2}$ shell
occupation by holes as spin increases to $\Imax$=25 in \up90Rh in
the SSSV calculation.

The calculated $E2$ transition strengths in \up90Rh are shown in
fig.\ref{fig-Rh90E2}$a$. The $B(E2)$ values calculated using SSSV
interaction are larger than those obtained in GF or SLGT0
calculations, implying larger deformation and thus greater
collectivity, as also seen from the energy plot with a rotational
reference subtracted, fig.\ref{fig-RhEnergies}$b$. Additional reason
for smaller $E2$ transition strengths in more restricted calculations
is that $p_{1/2}$ does not contribute, thus there is only contribution
from the $g_{9/2}$ shell. In the case of half-filled $j$-shells, there is
an additional suppression of the $E2$ transition strengths due to
so-called ``the center-of-the-shell selection rule''
\mcite{[p.191]}{Lawson}: in a single-$j$ configuration the
spectroscopic moment vanishes, and the $E2$ transition can only take
place between states differing in seniority by two units, if this
$j$-shell is half-filled. Indeed, the occupancies
(fig.\ref{fig-Rh90Occupancies}) show that the $g_{9/2}$ shell is
almost always half-filled, $\Qspec\approx0$, and $B(E2)$ are
relatively small (fig.\ref{fig-Rh90E2}) in GF as well as in SLGT0
calculation. The negative sign of $\Qspec$\ calculated with the SSSV
interaction (fig.\ref{fig-Rh90E2}$b$) suggests prolate deformation
along the band.

All three calculations of \up90Rh predict a backbending at spin
$\Imax-$6=19$^+$, as would be expected from the discussion of band
properties in a single-$j$ shell assuming only $JT$=01 pairing
interaction being important (subsection \ref{ssect-Selection}). In the
SSSV case, the backbending may be explained in terms of a
band-crossing. The bands are shown in fig.\ref{fig-RhEnergies}$b$
where the calculated states are connected if there is a strong
stretched $E2$ transition between them. As can be seen from the
figure, yrare band crosses yrast band (``$K^\pi$=5$^+$ band'') between
spins $I$=19 and 21. For the spins below 21$\hbar,$ the excited band
has lower particle occupation of the $g_{9/2}$ shell (approximately
0.55) as compared to the yrast band value, $\approx$0.68
(fig.\ref{fig-Rh90Occupancies}). States 21$_{1,2}^+$ have almost
identical occupation numbers. For higher spins the $g_{9/2}$ shell
occupation drops down to 0.5 in the yrast band, but equals 0.67 and
0.60 at states with spins 23$_2$ and 25$_2$, respectively. The $B(E2)$
values along the band with a lower occupation of the $g_{9/2}$ shell
are very similar to the GF and SLGT0 results (cf
fig.\ref{fig-Rh90E2}$a$), except for the transition
13$_2^+$$\rightarrow$11$_2^+$ that is stronger, $\sim$14.4~W.u. The
$\Qspec$ values are approximately constant along this band, $-$10
e$\cdot$fm$^2$, while at $I$=21 $\Qspec$ decreases to $-$38.0
e$\cdot$fm$^2$. The continuation of ``$K$=5$^+$ band'' has larger
intra-band $B(E2)$ values, than those calculated for the yrast states,
except for the transition starting at spin 25$_2^+$. The values are
1.4, 10.6 and 12.9 W.u.\ for transitions from state 25$_2^+$ to state
23$_2^+$, from 23$_2^+$ to 21$_2^+$, and from 21$_2^+$ to 19$_1^+$,
respectively. The spectroscopic quadrupole moment for the states
$I$=21$_2$-25$_2$ is varying around the value of $-$50 e$\cdot$fm$^2$.

The $B(E2)$ value calculated using SSSV interaction is reduced for the
transition $I_\mathrm{i}$$\rightarrow$($I_\mathrm{i}-$2), where
$I_\mathrm{i}$=$\Imax-$4=21$\hbar.$ A straightforward explanation
could be the fact that this transition connects two ``rotational
bands'' having somewhat different structure, seen either in terms
of the $g_{9/2}$ shell occupancy or the spectroscopic quadrupole
moment.  An alternative explanation could be the
``center-of-the-shell'' selection rule, mentioned above: the vanishing
$E2$ transition would indicate purity of the half-filled single-$j$
configuration \mcite{[p.324]}{Lawson}. In subsection
\ref{ssect-Selection} it was discussed that the seniorities of
states $\Imax-$6 and $\Imax-$4 are equal, thus the conditions for the
selection rule are favourable. However, the $\Qspec$ does not vanish
neither at 19$_1^+$ nor at 21$_1^+$ (see fig.\ref{fig-Rh90E2}$b$),
suggesting that the selection rule is not involved. While the
structure of nuclear wavefunctions of those states is very different,
and that is the reason for the reduced $B(E2)$ value.

A comparison of the predicted spin dependencies of the $B(E2)$ values
in \up90Rh (fig.\ref{fig-Rh90E2}) yields another question: whether the
reduction of $B(E2)$ at $I_\mathrm{i}$=21$_1^+$ is a feature of the
SSSV interaction.  The more restricted calculations do not predict
reduction. On the other hand, they do not predict bandcrossing as
a reason for the backbending. A similar situation occurs in \up48Cr
calculated using the modified surface-delta interaction
\cite{AJSABandTermination} (MSDI) in the configuration space
$(f_{7/2}p_{3/2})^8$, see fig.\ref{fig-Rh90E2}$a$. This interaction,
fitted for this nucleus, describes excitation energies of the yrast
states in \up48Cr a bit better than the KB3 interaction does. The
lowest bands predicted in \up48Cr by both KB3 and MSDI
interactions are somewhat similar to those predicted in \up90Rh by the
SSSV: In \up48Cr the yrast band continues above the backbending
preserving the nuclear structure; however, the band, terminating at
spin $\Imax$=16$^+$, cannot be traced back below the backbending.  The
$B(E2)$ values calculated in \up48Cr using MSDI are systematically
smaller, most probably due to a restricted model space. Nevertheless there
is a reduction of the $B(E2)$ value for the transition starting at
spin $I_\mathrm{i}$=$\Imax-$4=12 in the MSDI calculation. (It is not
clear why the $\Qspec$ vanishes at spin 12$\hbar$, see
fig.\ref{fig-Rh90E2}$b$. The selection rule would suggest, that it is
an indication of a rather pure $(f_{7/2})^8$ configuration, however,
the calculated probability of this configuration at this spin value is
only 69\%. But on the contrary to the selection rule prediction, the
increasing probability of $(f_{7/2})^8$ is followed by the decrease of
$\Qspec$ for spins 14 and 16.)  However, the relative decrease of
$B(E2; 12_1^+ \rightarrow 10_1^+)$ was not observed
\cite{CrExpData}. This decrease is also not predicted by the KB3
\cite{A48SMcalc}. Thus, having confronted the different predictions
for the \up90Rh and \up48Cr nuclei, a conclusion may be drawn that
only lifetime measurements or the calculation in a more extended model
space may suggest a solution to the puzzle whether there is a decrease
in $B(E2; I_\mathrm{i}
\rightarrow I_\mathrm{i}-2)$ values at $I_\mathrm{i}$=21 in \up90Rh.

\subsection{Estimation of the role of pairing interactions}
\label{ssect-Approach}

The role of {\em normalized\/} $L$=0 pairing interactions
\cite{DufourZuker} is discussed in present paper. Notations for
the forces used in the latter reference are adopted. Thus \Pbar{01}
(or simply 01 as an index, e.g.\ $E_{01}$) will stand for the
normalized isovector $J$=0, $S$=0, $T$=1 pairing interaction, and
\Pbar{10} (or 10) will stand for the normalized isoscalar $J$=1,
$S$=1, $T$=0 pairing interaction. The pairing strength parameters are
taken from eq.(C3) in ref.\cite{DufourZuker}. They are defined in
terms of a constant multiplied by a scaling parameter, $\hbar\omega\,$:
${\bar G}_{01}$=$-0.32\hbar\omega$ 
and 
${\bar G}_{10}$=$-0.51\hbar\omega$.
An expression for $\hbar\omega$ is taken
from \cite{DufloZuker99}, and the resulting strengths are given in
Table~\ref{tbl-GValues}.

\begin{table}[bt]
\begin{center}
\begin{minipage}[t]{7cm}
\caption{Strength parameters of normalized pairing
interactions used in present paper}
\vspace{0.1cm}
\renewcommand{\arraystretch}{1.2}  
\begin{tabular}{cccc}\hline
\label{tbl-GValues}
Pairing type & \up22Na & \up48Cr & \up90Rh \\
\hline
\Pbar{01}, ${\bar G}_{01}$ & -3.80 & -3.19 & -2.68 \\
\Pbar{10}, ${\bar G}_{10}$ & -6.05 & -5.09 & -4.27 \\
\hline
\end{tabular}
\end{minipage}
\end{center}
\end{table}

The pairing interaction influence (i) to the state energy, (ii) the
nuclear wavefunction and (iii) quadrupole properties will be
discussed. Investigation is performed in a perturbative way, by
comparing results obtained with the full Hamiltonian and the
Hamiltonians having reduced content of some pairing interaction
(either \Pbar{01} or \Pbar{10}). It will be shown that in some cases,
a competition between the pairing and the rest of the residual
interaction may be seen. (Cf.\ discussion about the interplay between the
quadrupole and the monopole parts of the Hamiltonian in ref.\cite{QuasiSU3}.)

The Hamiltonian without \Pbar{JT} pairing, $H_{JT}(0),$ is defined as a full
Hamiltonian, $H,$ from which the normalized pairing interaction,
$W_{JT},$ has been subtracted (in a similar way KB3-P01 and KB3-P10
were defined in ref.~\cite{PovesMartinezCr48}, but the pairing interactions
were not normalized. Note, however, that 
normalized and unnormalized forms are equivalent for a single major shell): %
\begin{equation}
	H_{JT}(0) = H - W_{JT},
\label{eq-NoPairingHam}
\end{equation}
where $JT$=01 for the isovector (\Pbar{01}) and 10 for the
isoscalar (\Pbar{10}) $L$=0 pairing interaction; and the meaning of 0 in
parentheses will be seen later. To investigate the role of a $JT$ pairing
interaction in a perturbative way, lets define a Hamiltonian
$H_{JT}(\lambda)$ that has some content of the pairing interaction
indicated by a weight parameter $\lambda$ (0$\leq$$\lambda$$\leq$1): %
\begin{equation}
	H_{JT}(\lambda) = H_{JT}(0) + \lambda\, W_{JT}.
\label{eq-SomePairingHam}
\end{equation}
Thus $H_{JT}(\lambda)$ ($\equiv$$H_{JT}$) is a linear interpolation
between the Hamiltonian without pairing, $H_{JT}(0)$, and the full
Hamiltonian, $H_{JT}(1)$=$H$. One can parameterize the eigenvalues of
this Hamiltonian in the following way:
\begin{equation}
	E_{JT}(\lambda) = E_{JT}(0) + \Wav{JT} \lambda + \chi(\lambda),
\label{eq-Energy}
\end{equation}
where $E_{JT}(0)$ is an eigenvalue of the Hamiltonian without
\Pbar{JT} pairing interaction, $E_{JT}(\lambda)$ is an eigenvalue of
the Hamiltonian $H_{JT}(\lambda),$ $\Wav{JT}$ is the expectation value
of the normalized $JT$ pairing interaction, and $\chi(\lambda)$
describes additional terms which are non-linear in $\lambda$. One can
see that $\chi$(0)$\equiv$0. If $\chi(\lambda)$ is set to 0, the
equation \refeq{eq-Energy} describes first-order perturbation to the
energy $E_{JT}(0)$. The energies are calculated and parameterized for
each spin value separately, i.e.\ the quantities defined by
eq.\refeq{eq-Energy} have additional index - spin $I$ - which is
omitted (thus the eigenvalues in full notation are
$E_{JT}(\lambda;I)$). The expectation value of the pairing
interaction, $\Wav{JT}$, may be evaluated directly from the calculated
nuclear wavefunction (cf.\
ref.\cite{MartinezLangankeNPPairing}). However, it was not possible to
perform such evaluation in present investigation.

\begin{figure}[tb]
\centerline{\psfig{figure=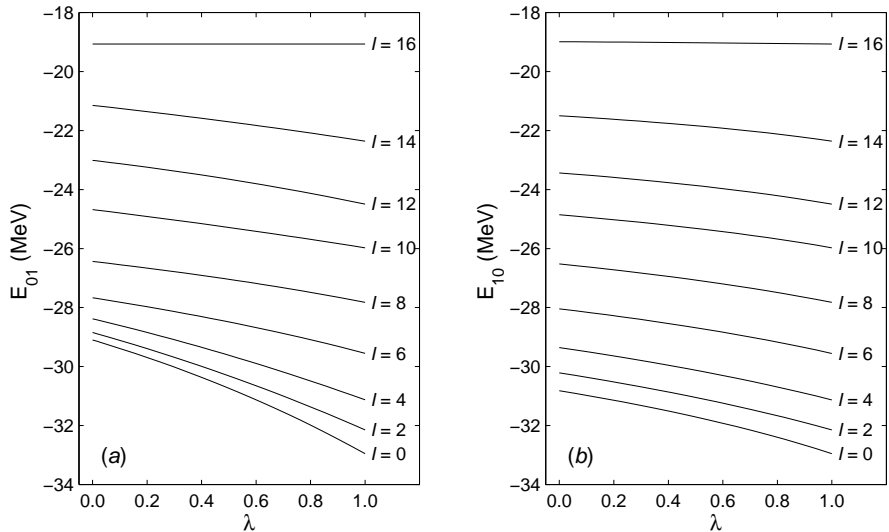,height=\figheight}}
\caption{%
Calculated energies of yrast states in \up48Cr as a function of the
\Pbar{JT} pairing interaction content. KB3 interaction is used. The
spin is indicated by a number close to the corresponding line. Plot $a$
($b$) shows the case $JT$=01 (10).}
\label{fig-Cr48ELambda}
\end{figure}

An example of the $H_{JT}$ eigenvalue spectrum dependence on the
pairing interaction content is given in fig.\ref{fig-Cr48ELambda},
where energies of the yrast states in \up48Cr calculated using the KB3
interaction are shown as a function of parameter $\lambda$. The case
when \Pbar{01} interaction content is varied is shown on the left,
while the \Pbar{10} case is shown on the right. The meaning of
parameters in eq.\refeq{eq-Energy} may be seen from this figure:
$\Wav{JT}$ is the slope of the $E_{JT}(\lambda)$ line at $\lambda$=0,
and $\chi(\lambda)$ is the deviation of $E_{JT}(\lambda)$ values from
the linear behaviour ($E_{JT}(0)$+$\Wav{JT}\lambda$). One can see that
$|\chi(\lambda)|$ increases together with $\lambda$ (the lines bend
down), and that non-linear contributions are largest for lower
spins. Since the slope of the $E_{JT}(\lambda)$ line is proportional
to the average number of $JT$-coupled pairs, it is clearly seen that
there are more such pairs in the nuclear wavefunction calculated with
a Hamiltonian close to the full Hamiltonian. Which may be understood
from the fact that pairing interactions increase the binding energy of
a pair, thus pairing correlations are more favourable if pairing force
is stronger, i.e.\ the pairing interaction content in the Hamiltonian
is larger.

Lets define the pairing energy, $\EpJT{JT},$ which will show the
increase in the binding energy of a state having spin $I$ due to an
addition of the $JT$ pairing interaction:
\begin{equation}
	\EpJT{JT} \equiv E_{JT}(0)-E(1),
\label{eq-PairingEnergy}
\end{equation}
where $E(1)$ is the state energy calculated with the full Hamiltonian,
and $E_{JT}(0)$ is the state energy calculated using the Hamiltonian
without \Pbar{JT} pairing interaction; additional index - spin $I$ - is
omitted. The pairing energy will always be non-negative, since the
pairing interactions are attractive, i.e.\ they increase binding.

A Hamiltonian may be considered as consisting of monopole and
multipole parts~\cite{DufourZuker}, where the monopole part is
responsible for the correct nuclear binding energies at the observed
nuclear radii and for the single-particle behaviour, and the multipole
(or residual) part describes collective properties and is responsible
for configuration mixing. One may notice that the pairing interactions
contribute to both parts: the changed content of the pairing
interaction affects the monopole part due to the moved centroids of
the total interaction, while the multipole part is affected because
this interaction is responsible for correlations (only this part of a
pairing interaction, that contributes to the multipole Hamiltonian, may be
viewed as being responsible for pair correlations in the spherical
mean-field sense). The strengths of pairing interactions were derived
in ref.\cite{DufourZuker} from a monopole-free Hamiltonian. However,
in present investigation it was not possible to make exact separation
of the pairing effects into monopole and multipole parts in a
consistent way, thus only total effect is investigated. One term in
the monopole part of the Hamiltonian, namely, the single-particle
contribution, may be readily evaluated having $j$-shell occupation
numbers. The single-particle term may be written as:
\begin{equation}
	\spt=\sum_{\{j\}} n_j \varepsilon_j,
  \label{eq-KineticTerm}
\end{equation}
where $\varepsilon_j$ is the single-particle (or single-hole) energy
of a $j$-shell, and $n_j$ is the average number of particles in the
shell.

In a similar manner as the change of binding energies (the pairing
energy) is defined in eq.\refeq{eq-PairingEnergy}, one can define the
change of the single-particle contribution as an outcome of pairing
interaction:
\begin{equation}
	\Delta \spt_{JT} = \spt_{JT}(0)-\spt(1),
\label{eq-DeltaKd}
\end{equation}
where $\spt_{JT}(0)$ and $\spt(1)$ is the single-particle term
evaluated using occupation numbers calculated with the Hamiltonian
without \Pbar{JT} pairing and with the full Hamiltonian,
respectively. This difference may give some idea about the change in
the mixing with the high-lying $j$-shells as a result of pairing
interaction, because particles occupying shells with large
single-particle energy $\varepsilon_j$ make larger influence to the
total single-particle contribution, see
eq.\refeq{eq-KineticTerm}. (For example, change in the occupation of
the $f_{5/2}$ shell by 0.1 particle and corresponding change in the
$f_{7/2}$ shell occupation yields 0.6 MeV in $\Delta \spt,$ if the KB3
interaction is used.) In addition, it gives an idea of a fraction of
the pairing energy coming due to single-particle contributions.

The difference $\Delta \spt_{JT}$ can be both positive and
negative. If it is negative, more particles are excited to higher
$j$-shells in the full calculation, while the pairing interaction
(over)compensates the loss in the single particle contribution. If it
is positive, the pairing interaction contributes to the gain of energy
by favouring configuration with less particles excited from the lowest
$j$-shells.

\section{Results and discussion}
\label{sect-Results}

It is known that the eigenfunctions of a well-bound states are less
sensitive to a slight change of the potential, than the corresponding
eigenvalues are. If two shell model Hamiltonians have two-body
interaction being different by content of a pairing force, their
eigenvalues are different but nuclear properties calculated using the
corresponding eigenfunctions may be similar. The discussion of the
role of pairing interaction is started from investigation of this
interaction's influence on the nuclear wavefunction and quadrupole
properties (subsection \ref{ssect-Wavefunction}). It is known, that
the pairing interaction does not affect quadrupole properties
\cite{QuasiSU3,PovesTalk}. It is, however, shown here that there are
cases when this is not true. The influence of pairing interactions on
energies is discussed afterwards, in subsection
\ref{ssect-StateEnergy}. A strict separation into these two topics
cannot be made, thus the energies are also discussed together with
wavefunctions.

			\clearpage
\centerline{\psfig{figure=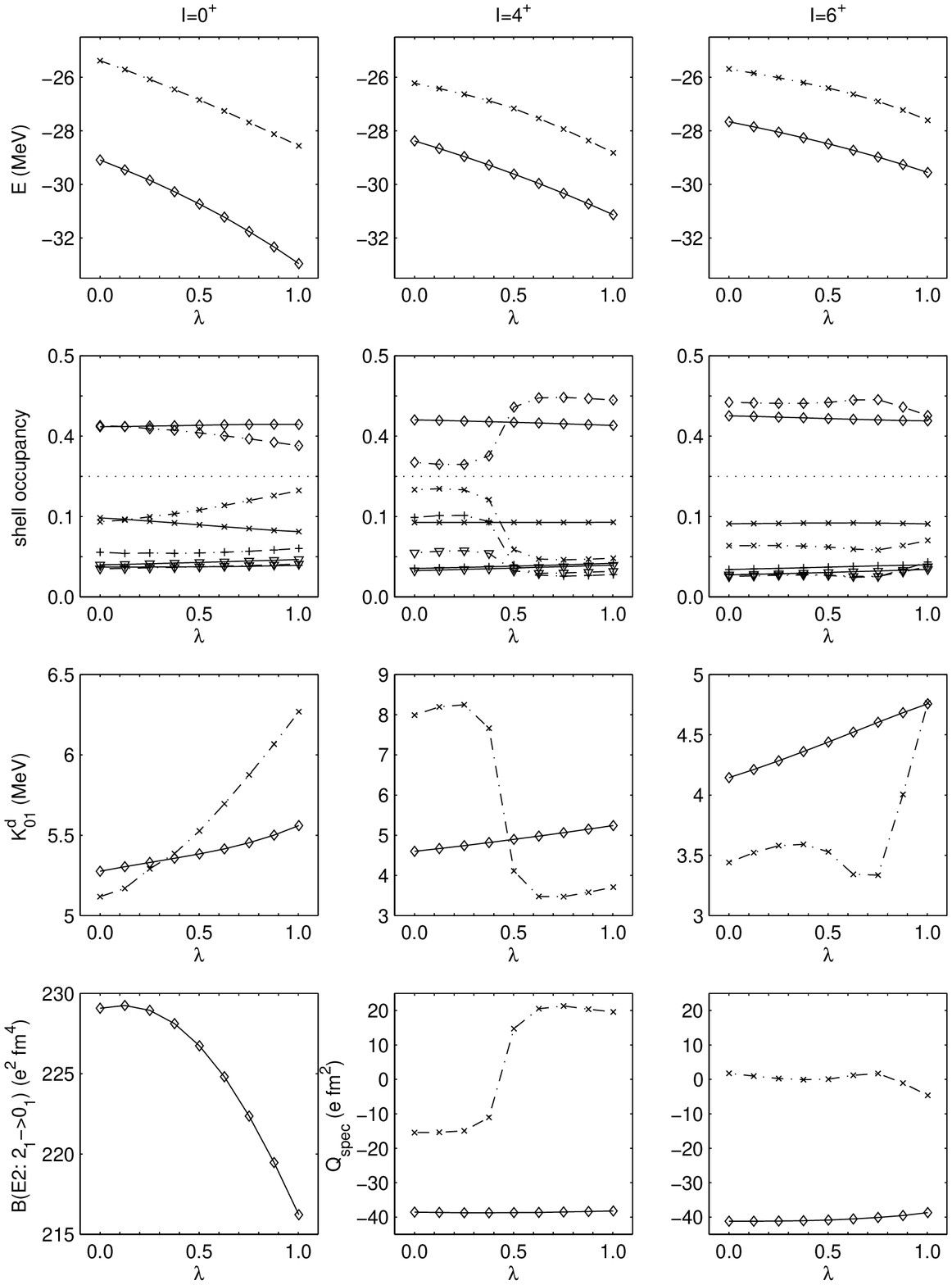,width=12cm}}
\begin{figure}[h]
\caption{
Energies, occupancies of the shells, single-particle contributions,
and $E2$ properties of yrast and yrare states in \up48Cr calculated at
spins $I^\pi$=0$^+,$ 4$^+$ and 6$^+$ using the KB3 interaction. The
\Pbar{01} pairing interaction content is given on the horizontal axis
($\lambda$=0 means no-pairing interaction, while $\lambda$=1 means
full pairing interaction is present in the Hamiltonian). The top-row
plots show the energies, $E$; the second-row plots show shell
occupancies; the third-row plots show the single-particle
contributions, $\spt_{01}$; and the bottom-row plots show $E2$
properties: $E2$ transition strength between yrast 2$^+$ and 0$^+$
states, and the spectroscopic quadrupole moment $\Qspec$ at spins
$I$=4$^+$ and 6$^+$. The solid and dot-dashed lines are used to
distinguish the yrast and yrare states, respectively. The $f_{7/2}$
shell occupancy is marked by diamonds, $p_{3/2}$ by crosses, $p_{1/2}$
by pluses, and $f_{5/2}$ by triangles.  
}
\label{fig-Cr48I6Occupancies}
\end{figure}

\subsection{Nuclear wavefunction and quadrupole properties}
\label{ssect-Wavefunction}

In general, the change in energy of a state $E_{JT}(\lambda;I)$ as
pairing content (i.e.\ the $\lambda$ value) increases, illustrates two
effects. Besides a direct effect coming from the increased binding of
certain nucleon pairs (linear in $\lambda$), there is an additional
increase in energy (non-linear) due to the change of the number of
such pairs that also results in abandoning of some less favourable
configurations, i.e.\ the ``structural effect'' or $\chi$ in
eq.\refeq{eq-Energy}. From fig.\ref{fig-Cr48ELambda} one may see that
the structural effect increases the binding energy: if we start from
the Hamiltonian without, say, \Pbar{01} pairing, there are not so many
$JT$=01 coupled pairs, and the slope is not large. But as $\lambda$
approaches 1, it becomes more and more favourable to form such pairs,
and the $E(\lambda;I)$ curve bends down. If we start from the full
Hamiltonian and reduce the amount of \Pbar{01} pairing, the energy
loss would be much larger if the nuclear wave function is fixed than
it is when configuration change is allowed.

Investigation of the pairing interaction effect on a particular state
in a perturbative way may reveal a gradual change of the state
properties. Changes are larger, when there is another close lying
state, since these states may mix in a different proportion following
the change of the interaction. As a result, when varying $\lambda,$
one can see a mutual interchange of state properties. In an extreme
situation, either sharp or gradual transition of the state properties
from those of one state to those of the other state may be observed.

Let us first investigate a case, when there is a substantial energy
distance between two states, thus other states do not influence the
change of the state properties. The fig.\ref{fig-Cr48I6Occupancies}
shows some properties of the first and second excited states in
\up48Cr calculated at spins $I$=0, 4 and 6$^+$ using the KB3
interaction, and the content of \Pbar{01} interaction is varied. Those
properties are: state energy, $j$-shell occupation numbers,
single-particle contribution to the total energy, and spectroscopic
quadrupole moment $\Qspec$ (or $E2$ transition strength where $\Qspec$
is not defined). It is seen that the energy of the yrast states
decreases (binding increases) as $\lambda$ increases, but the
occupancies almost do not change (except for the $p_{3/2}$ shell
occupancy at 0$^+$). Some occupancies are difficult to distinguish in
the plot, however, their values are not important for the discussion.
The occupancies may be summarized in terms of the single particle
contribution to the state energy, see eq.\refeq{eq-KineticTerm}. As
can be seen from the figure, $\spt_{01}$ of the yrast states changes
only slightly or not at all. One may also notice that the wavefunction
of the yrare state at 0$^+$ is more affected by the change in pairing
force than the yrast state (occupation numbers and thus $\spt_{01}$
change more than in the yrast state).
In addition, the second state at $I$=4$^+$
interacts with the third state, and the occupancies, the
single-particle contribution as well as spectroscopic quadrupole
moment change drastically at $\lambda$$\approx$0.45, while the state
energy only slightly changes slope. A similar change is seen at
$6_2^+$.

The bottom-row plots in fig.\ref{fig-Cr48I6Occupancies} show
calculated $E2$ properties in \up48Cr as a function of \Pbar{01}
pairing interaction content. As may be seen, the behaviour of the
spectroscopic moment, $\Qspec$, is very similar to that of the
single-particle contribution, $\spt_{01}$: if $\spt_{01}$ does not
change, the $\Qspec$ does the same. However, it is wrong to correlate
the increasing $f_{7/2}$ shell occupancy ($\spt_{01}$ tends to 0) with
the more oblate nuclear shape ($\Qspec$$>$0). The
``center-of-the-shell'' rule (mentioned in subsection
\ref{ssect-Experiment} above) demands $\Qspec$=0 for $\spt$=0,
however, as can be seen from the figure, small admixtures of other
$j$-shells causes deviation from the rule. The bottom-left plot shows
the strength of the stretched $E2$ transition between the lowest yrast
states, $2_1^+$$\rightarrow$$0_1^+$. As more pairing interaction is
present in the Hamiltonian ($\lambda$ increases), less particles are
excited from the $f_{7/2}$ shell ($\spt_{01}$ decreases), thus prolate
nuclear deformation decreases (cf.\ the behaviour of $\spt_{01}$ and
$\Qspec$ at $I$=$4_2^+$). This is in agreement with the
``center-of-the-shell'' rule.  However, the low probability of pure
$f_{7/2}^8$ configuration (see below) as well as changing $\Qspec$ of
$2_1^+$ state (not shown here) suggests that a bit more sophisticated
picture than the mentioned selection rule is needed. Note, however,
that the reduction of the $B(E2)$ value by 6\% means only
insignificant change in an approximate quadrupole deformation
$\beta_2$ (calculated assuming the axially symmetric rotor model is
valid): from $\beta_2$=0.270 to 0.263.

\begin{figure}[bt]
\centerline{\psfig{figure=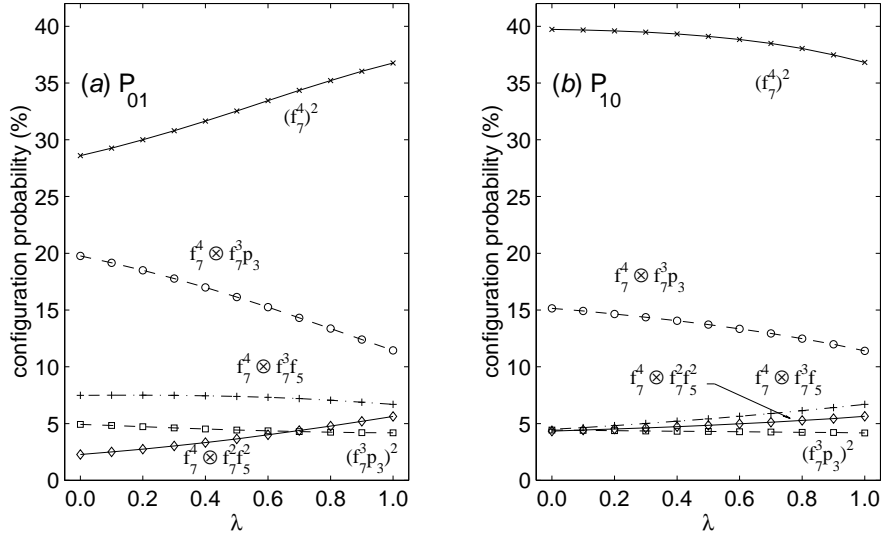,height=\figheight}}
\caption{%
Calculated probabilities of configurations in the \up48Cr ground state
as a function of \Pbar{JT} pairing interaction content, $\lambda$:
the left and right plots show \Pbar{01} and \Pbar{10} cases, respectively.
The KB3 interaction was used in the calculation.
Only configurations with probability exceeding 5\% at some point are
shown. The missing probability is about 33\%.  }
\label{fig-Cr48I0Wf}
\end{figure}

Although the binding energy of the yrast state changes significantly
due to a change in the pairing interaction content, the occupation
numbers change only slightly, and only a more detailed view of the
probabilities of different configurations reveal changes. As an
example,
configurations in the ground state of \up48Cr are shown in
fig.\ref{fig-Cr48I0Wf}. Those configurations which never achieved
probability larger than 5\% are omitted. The missing probability,
i.e.\ summed 
probability of not shown configurations, constitutes 35\%
and 31-35\% for \Pbar{01} and \Pbar{10} cases, respectively. The
average numbers of particles in $j$-shells are unevenly distributed
among the omitted configurations. For example, the missing particle
numbers are 1.8 for $f_{7/2}$, 0.5 for $p_{3/2}$, 0.2 for $p_{1/2}$
and 0.4 for $f_{5/2}$ shell in the case of \Pbar{01} pairing
interaction. The probabilities marked like
$f_{7/2}^4$$\otimes$$f_{7/2}^3p_{3/2}^{}$ are added probabilities of
configurations $\pi f_{7/2}^4\nu (f_{7/2}^3p_{3/2}^{})$ and $\pi
(f_{7/2}^3p_{3/2}^{}) \nu f_{7/2}^4$ (here $\pi$ stands for protons
and $\nu$ for neutrons), while marks like $(f_{7/2}^4)^2$ indicate
that protons and neutrons have the same configuration ($f_{7/2}^4$ in
this example). The fig.\ref{fig-Cr48I0Wf}$a$ shows that the
probability of configurations where all 8 particles are in the
$f_{7/2}$ shell is increasing by 8\% in the \Pbar{01} case, when
$\lambda$ goes from 0 (no pairing) to 1 (full pairing). At the same
time, the configurations with one particle excited to $p_{3/2}$
decreases by the same amount. The \Pbar{01} pairing increases the
probability of configurations with two particles of the same kind
being excited to $f_{5/2}$ by 3\%, but this probability reaches only
about 6\% at the maximum. The mixing with $p_{3/2}$ may be favourable
at $\lambda$=0, but at $\lambda$=1, where $JT$=01 pairs are more
favoured in energy, shells with large $\ell,$ i.e.\ $f_{7/2}$ and
$f_{5/2}$, are more important. The fig.\ref{fig-Cr48I0Wf}$b$ shows
that the \Pbar{10} interaction lowers the probability of $f_{7/2}^8$
as well as $f_{7/2}^7p_{3/2}^{}$ configurations by 4\% each; while the
probabilities of configurations with one or two particles in the
$f_{5/2}$ shell slightly increase.

Information about changes in the nuclear wavefunction are still hidden
in fig.\ref{fig-Cr48I0Wf}: the linear change of configuration
probabilities does not explain non-linear change of the state energy
(see fig.\ref{fig-Cr48ELambda}). It may be suspected that the average
seniority changes in a nonlinear manner.

An example, where the two lowest-lying states affect each other when
the pairing interaction content changes, may be taken from \up90Rh
calculated using the SSSV interaction.
Fig.\ref{fig-Rh90I17Occupancies} illustrates three cases of the two
interacting states calculated for the \Pbar{01} force: weak mutual
influence at spin $I^\pi$=13$^+$, smooth ``crossing'' at 17$^+$, and
sharp crossing at 19$^+$. The states hardly influence each other at
spin $I$=13$^+$ (see left column plots in
fig.\ref{fig-Rh90I17Occupancies}). Although the energy difference
between the states remains approximately constant as more pairing is
added (i.e., $\lambda$ approaches 1), properties of the states change
in a correlated way: in the yrast state the $g_{9/2}$ shell becomes
more mixed with other $j$-shells, while the opposite is seen in the
yrare state. This change in shell mixing can be seen both from the
shell occupancies by particles and from the single-hole contribution,
$\spt_{01}$. Note that the single-hole contribution in \up90Rh behaves
in an opposite way than the single-particle contribution in \up48Cr
when trend is compared to the occupation of the high-$j$ shell: if the
$f_{7/2}$ shell occupancy increases, cf.\
fig.\ref{fig-Cr48I6Occupancies}, the single-particle contribution in
\up48Cr decreases, while the $g_{9/2}$ occupancy by particles and the
single-hole contribution in \up90Rh increase or decrease together. The
two lowest states come closer in energy and separate again at spin
$I$=17$^+$ (see middle column plots in
fig.\ref{fig-Rh90I17Occupancies}). The behaviour of occupation numbers
(or of the single-hole contribution) suggests that the states
gradually exchange their properties, since the shell occupancies of
one state at $\lambda$=0 become equal to those of the other state at
$\lambda$=1. A similar picture is seen at spin $I$=19$^+$ where the
two states approach and cross each other without much interaction: the
energy slope changes sharply at $\lambda$=0.9 and both the occupancies
and thus the single-hole contributions change abruptly. In this case
it is possible to calculate the pairing energy of a specific
configuration, i.e.\ of the configuration being yrast at $\lambda$=1,
or of the one being yrast at $\lambda$=0.

The bottom-row plots of fig.\ref{fig-Rh90I17Occupancies} show $\Qspec$
calculated in \up90Rh using the SSSV interaction for the three spin
values discussed above. It is clearly seen that $\Qspec$ is correlated
with the single-hole contribution, $\spt_{01}$. For the less mixed
configurations (small $\spt_{01}$ values), $\Qspec$ tends to zero;
while for the more mixed configurations (larger $\spt_{01}$), the
prolate deformation increases ($\Qspec$ decreases). A picture of the
mutual influence between the two states, obtained plotting the deformation
coordinate (or $\Qspec$, assuming the axial symmetry) at horizontal
axis and energy at vertical axis, would be somewhat different than the
one described above. If the positions of points representing values 
of
first and second states for a given spin are followed as $\lambda$
goes from 0 to 1, one would see, that the ``minima'' are well
separated by deformation, and they merely ``sink'' (i.e., the binding
increases) at spin 13$^+$. A bit more complicated picture occurs at
spin 17$^+$, where minima are separated for $\lambda$ values close to
the limits, but at $\lambda$$\approx$0.7 the two states should be
treated as first and second state in the same potential well and not
as separate minima.  The properties of states are also very similar
there (see fig.\ref{fig-Rh90I17Occupancies}, middle column). At spin
19$^+$, there minima are well separated by deformation for
$\lambda$=0. But for larger $\lambda$ values, the well depths increase
at a different rate and become more and more  
similar. As a result,
the wavefunctions start to mix, and finally, the state having larger
deformation becomes yrast. A similar sharp transition between two
states may occur in the backbending regions, if the crossing bands
have different pairing properties.
\centerline{\psfig{figure=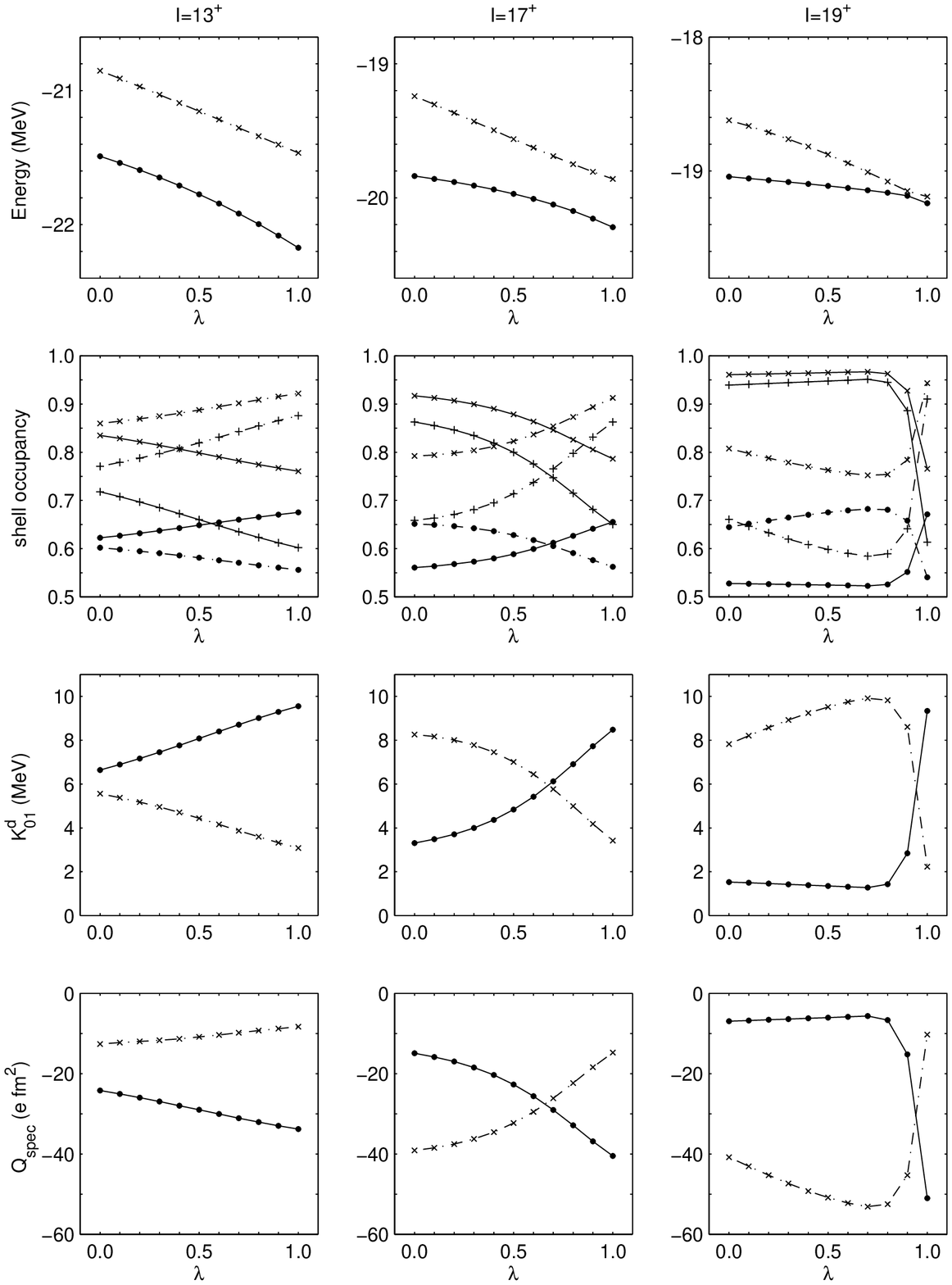,width=12cm}}
\vspace{-1.0cm}
\begin{figure}[h]
\caption{%
Some properties of yrast and yrare states in \up90Rh calculated at
spins $I$=13$^+$, 17$^+$ and 19$^+$, using the SSSV interaction, as a
function of the \Pbar{01} pairing interaction content. For explanations
see fig.\ref{fig-Cr48I6Occupancies}. The particle occupancy of the $g_{9/2}$
shell is shown by filled circles, $p_{3/2}$ by crosses, $p_{1/2}$ by
pluses. The single-hole contribution $\spt_{01}$ is shown in the third-row 
plots.}
\label{fig-Rh90I17Occupancies}
\end{figure}

From the presented here cases of two interacting states one may see,
that the ($JT$=01) pairing increases shell mixing and may unfavour the
pure high-$j$ shell configurations. Thus deviations from the
single-$j$ picture of the pairing energy spin dependence, discussed in
subsection \ref{ssect-Selection}, may be expected. Although only the
\Pbar{01} cases were discussed here, a similar picture is obtained
when \Pbar{10} interaction content is varied.

\subsection{State energy}
\label{ssect-StateEnergy}

In the previous subsection it was already discussed that the increase
of pairing interaction content yields additional binding of the
state. This subsection discusses changes in the binding energy of
yrast states. First of all, investigation of the obtained pairing
energy dependence on the model space is presented. Then pairing energy
dependence on the residual interaction is briefly reviewed. And
finally, the pairing energy in the selected bands is discussed.

\subsubsection{Model space effects}
\label{ssect-ModelSpace}

During discussion of the role of pairing interactions, a natural
question may be asked, whether the model space is sufficient to trust
the results. For example, the definition of schematic pairing forces
($L$=0) suggests that it is important to include both the high-$j$
shell and its spin-orbit partner in the model space, in order to get
reasonable pairing energy values. Thus the model space for \up90Rh is
probably too small, since the $g_{7/2}$ shell is excluded. This
section tries to shed some light on this problem. 

\begin{figure}[tb]
\centerline{\psfig{figure=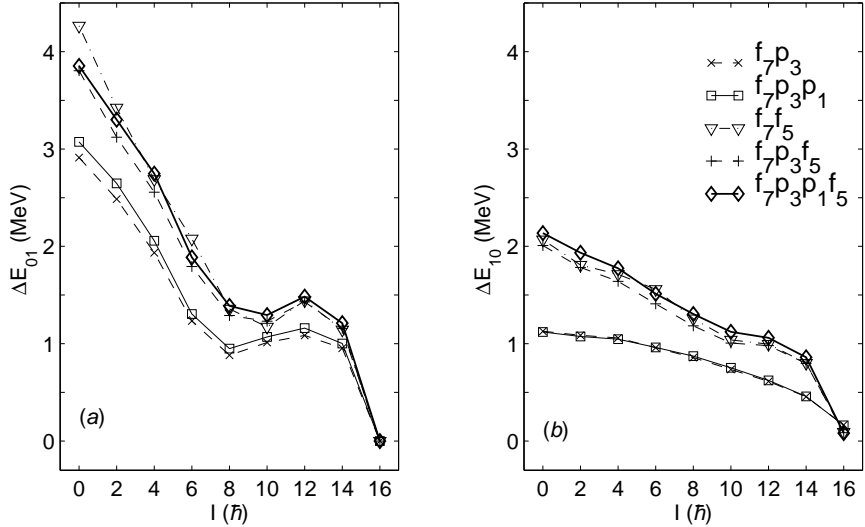,height=\figheight}}
\caption{
Model space dependence of the pairing energy. Calculated for the
\up48Cr nucleus using KB3 interaction.  The left (right) plot
shows \Pbar{01} (\Pbar{10}) case.}
\label{fig-EpCr48Investigation}
\end{figure}

To investigate the question whether the pairing energy, defined by
eq.\refeq{eq-PairingEnergy}, is model space independent, yrast states
of \up48Cr were calculated in different model spaces made of different
combinations of $j$-shells from the $pf$ major shell but using the
same residual interaction, KB3. Experimental energies of the yrast
states are reproduced with a different success in the selected model
spaces, but in this investigation only the response to the change of
pairing interaction content is important. The derived pairing energies
(defined as energy difference between yrast states calculated with the
Hamiltonian having no \Pbar{JT} pairing interaction and with the full
Hamiltonian) are shown in fig.\ref{fig-EpCr48Investigation}:
\Pbar{01} case is shown on the left, and \Pbar{10} case is shown on the
right. It may be seen that the largest contribution comes from the
spin-orbit partners having large orbital angular momentum $\ell$. Thus
the change in binding energies calculated in the full $pf$ model space
is almost completely accounted for in a smaller model space containing
both $f_{7/2}$ and $f_{5/2}.$ It is somewhat surprising that there are
no obvious signs of a model space increase from $f_{7/2}f_{5/2}$ to
full $pf$ for both \Pbar{01} and \Pbar{10} forces. This may be
understood taking into account the relatively small orbital angular
momentum of the $p_{3/2}$ and $p_{1/2}$ shells, as well as from small
occupancies of the $j$-shells other than the high-$j$ shell (see
fig.\ref{fig-Cr48I6Occupancies} and ref.\cite{SMCHFBCr48}).

The similarity of the pairing energy in the $f_{7/2}f_{5/2}$ and full
$pf$ calculations suggests that there is some saturation of the
pairing interaction influence, and the change in nucleus structure
accounts for the interaction change. That is also suggested by the model
space dependence of the change in the single particle contributions (see
below).

One conclusion which could be drawn from this investigation is that
the approximate energy change due to the \Pbar{01} force as a function
of spin can be obtained from the calculation in a model space without
the spin-orbit partner, for example, $f_{7/2}p_{3/2}$ or
$f_{7/2}p_{3/2}p_{1/2}$ (see
fig.\ref{fig-EpCr48Investigation}$a$). The absolute magnitude can be
estimated multiplying the pairing energy obtained in a more restricted
model space by the ratio of model space degeneracies. For example, if
one takes the result of the $f_{7/2}p_{3/2}p_{1/2}$ model space and
multiplies it by 20/14, the estimated $\EpJT{01}$ for the full $pf$
calculation is only slightly too large: the difference is about 0.3
MeV for spins $I$=0 and 2$^+$ and less for other spin values.
However, the predicted slope between states having spins $I$=8 and 10
is different from that in the full calculation. Such
scaling does not work for the \Pbar{10} force (see
fig.\ref{fig-EpCr48Investigation}$b$), where inclusion of $f_{5/2}$ is
necessary to obtain the profile of the $\EpJT{10}$ spin-dependence. One
may also notice, that the inclusion of the $f_{5/2}$ shell into the
model space approximately doubled the calculated \Pbar{10} pairing
energy.

This investigation, however, leaves the question, whether inclusion of
the other major shells would change the pairing energy.  One would
expect that the influence of higher lying $j$-shells is not large
because a large difference in single-particle energies would limit the
mixing. If, however, the shells below the lowest partially-filled
shell (the ``lower shells'') are included in the model space, the
pairing interaction effect should be separated into a part coming from
the lower shells and the rest, because the lower shells always react
to the change of the pairing interaction content (since they are
mostly filled).

\begin{figure}[tb]
\centerline{\psfig{figure=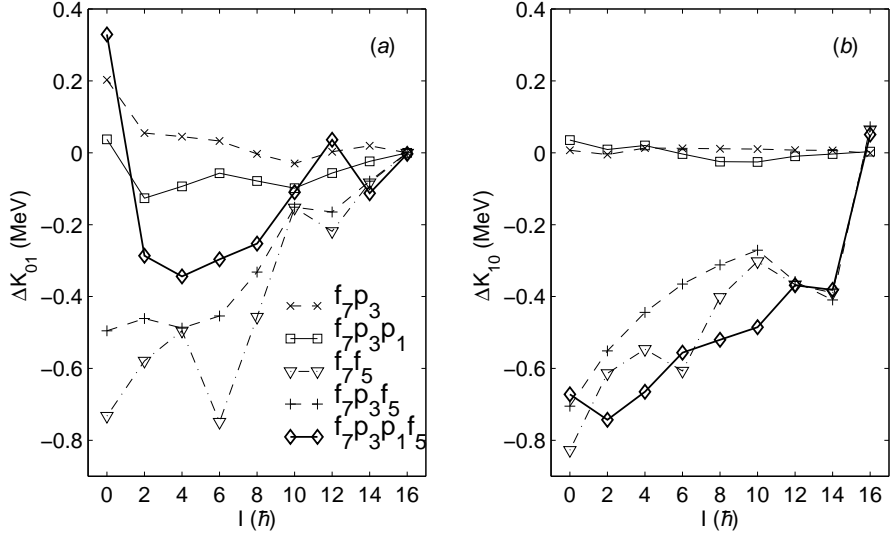,height=\figheight}}
\caption{Model space dependence of the difference in
single-particle contribution, $\Delta\sptJT{JT}$=$\sptJT{JT}(0)$$-$$\sptJT{JT}(1)$.
Calculated for nucleus \up48Cr using the KB3 interaction.
The left (right) plot shows \Pbar{01} (\Pbar{10}) case.}
\label{fig-KdCr48Investigation}
\end{figure}

The model space dependence of a difference in the single-particle
contribution, $\Delta\sptJT{JT},$ defined by eq.\refeq{eq-DeltaKd}, is
shown in fig.\ref{fig-KdCr48Investigation}. This difference gives some
idea about the change in the mixing with the high-lying $j$-shells as a
result of pairing interaction. It can be seen that $\Delta\sptJT{JT}$
behaves in a rather irregular way, as compared to the total change in
energy (fig.\ref{fig-EpCr48Investigation}), and it is not possible to
predict how the contribution will change with an increase of the model
space. However, there are also similarities in the change of the
single-particle contributions. For example, it is clearly seen, that
the pairing interactions favour configurations with more particles
excited to higher shells in the model spaces containing both
$f_{7/2}$ and $f_{5/2}$ shells (except for $I$=0$^+$ and 12$^+$ states
calculated for the \Pbar{01} case in the full $pf$ model space), since
$\Delta\sptJT{JT}$$<$0. The opposite is seen for most spin values in
the $f_{7/2}p_{3/2}$ \Pbar{01} case. It is also interesting to note
that the \Pbar{10} force decreases shell mixing at the maximum spin
state $\Imax$=16$^+,$ since $\Delta\sptJT{JT}$ is positive here.

In summary, the investigation of the model space dependence of the
pairing energy suggests that the model space containing shells
slightly above the lowest partially-filled shell is sufficient to
investigate the influence of the \Pbar{01} pairing interaction on the
state energies, while the \Pbar{10} pairing energy can be obtained
only in model spaces containing both the high-$j$ shell and its
spin-orbit partner.

\subsubsection{Residual interaction dependence}
\label{ssect-ResidualInteraction}

Somewhat different investigation is the calculation of energy changes
in the same nucleus using different interactions.  If the residual
force is unique, the difference in the calculated pairing energy would
come only from the model space difference. However, since there is no
unique force yet, the agreement in the pairing energy, if occurs, is
more likely to be accidental, and there should be clear dependence on
the chosen residual interaction. A separate question is whether this
dependence should show up {\em only\/} as different nonlinear effects,
$\chi$ in eq.\refeq{eq-Energy}. If the nuclear mean-field
(Hartree-Fock field) is universal (cf.\ discussion about the universal
multipole part of the Hamiltonian in ref.\cite{DufourZuker}, that gave
foundation for this investigation), one would expect that the average
number of pairs in the wavefunction calculated using the Hamiltonian
without pairing interaction is independent on the chosen (full)
residual interaction. A disagreement in this average number of pairs
would come from the other multipole terms favouring certain
two-nucleon states, different in different residual interactions. The
average number of pairs is proportional to $\Wav{JT}$ in
eq.\refeq{eq-Energy}, thus more information will be provided in the 
subsection \ref{ssect-PairingEnergy}, when the coefficients of the
state energy decomposition, eq.\refeq{eq-Energy}, will be presented.

\begin{figure}[tb]
\centerline{\psfig{figure=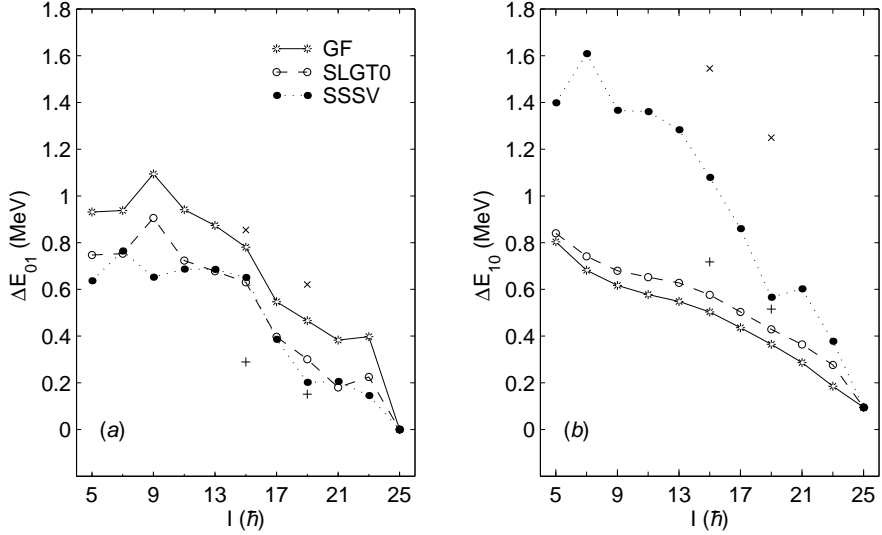,height=\figheight}}
\caption{Interaction dependence of the pairing energy
calculated for \up90Rh nucleus using three interactions.
The left (right) plot shows \Pbar{01} (\Pbar{10}) case.
The GF and SLGT0 values are shifted down to match the SSSV value at spin
$I$=25$^+.$ For spins $I$=15$^+$ and 19$^+,$ the SSSV energy differences 
between yrare at $\lambda$=0 and yrast at $\lambda$=1 states are indicated
by cross signs, and the energy differences between states behaving in an
opposite way are marked by plus signs.
}
\label{fig-EpRh90Investigation}
\end{figure}

The investigation was performed for \up90Rh using GF and SLGT0
interactions in the configuration space $(p_{1/2}g_{9/2})^{14}$, as
well as the SSSV interaction in $(g_{9/2}p_{1/2}p_{3/2})^{-10}.$ The
calculated pairing energies are shown in
fig.\ref{fig-EpRh90Investigation}. To take into account the effect of
the lower shells (in our case, of the $p_{1/2}$ shell), which was
discussed in subsection \ref{ssect-ModelSpace}, pairing energy
calculated in \up90Rh using GF and SLGT0 interactions was
approximately corrected taking the maximum spin state $\Imax$=25$^+$
as a reference. There spin is produced by 10 particles in $g_{9/2}$
shell, and the $p_{1/2}$ shell is full. There is no non-linear
dependence on the pairing interaction content, i.e.\
$\chi(\lambda)$$\equiv$0 at this state. Thus the values obtained in
the GF and SLGT0 calculations are shifted down to match the SSSV value
at the maximum spin state $\Imax$=25$^+$. GF pairing energies are
reduced by 0.66 and 0.24 MeV for \Pbar{01} and \Pbar{10} cases
respectively. The corresponding values for the SLGT0 interaction are
0.80 and 0.12 MeV. As explained in the subsection
\ref{ssect-Wavefunction} above, due to a sharp ``crossing'' in the energies
$E_{JT}(\lambda;I)$ calculated with the SSSV interaction, it is
possible to calculate the pairing energy of configurations becoming
yrast in the full Hamiltonian and being yrare in the Hamiltonian
without pairing. These values, $E_2(0)-$$E_1(1)$ (here subscript
refers to the state number for a given spin and total isospin),
calculated at spins $I$=15$^+$ and 19$^+$ are marked by crosses in the
fig.\ref{fig-EpRh90Investigation}. The pairing energy of
configurations which become yrare in the full Hamiltonian,
$E_1(0)-$$E_2(1)$, are shown by pluses. The figure clearly shows that
the configuration becoming yrast in the calculation with the full
interaction gets much more pairing energy than the yrare configuration.

As seen from the figure \ref{fig-EpRh90Investigation}$a$, the spin
dependence of \Pbar{01} pairing energies, obtained from different
interactions, is very similar, especially between the GF and SLGT0
interactions calculated using the same model space. However, the
absolute magnitude of pairing energy is different. The difference in
the calculated \Pbar{01} pairing energies using GF and SSSV
interactions cannot be explained by the model space effects (cf.\
fig.\ref{fig-EpCr48Investigation}): $\EpJT{01}$ from the SSSV
calculation should be larger than values obtained from the GF
calculation.  Thus it is the expected dependence on the residual
interaction.

The origin of the difference in the spin dependence of \Pbar{01}
pairing energy close to the band termination is not clear: in the SSSV
case, there is a small increase in pairing energy at spin 21$^+$ (in a
similar way as in \up48Cr at 
$\Imax-$4=12, see
fig.\ref{fig-EpCr48Investigation}), while in more restricted
calculations using GF and SLGT0 interactions, the \Pbar{01} pairing
energy is locally smaller at 21$^+$ and increases at $I$=23$^+$. 

The above discussion of the model space effects (subsection
\ref{ssect-ModelSpace}) suggested, that the calculated \Pbar{10}
pairing energies should be smooth if model space does not contain the
spin-orbit partner of the high-$j$ shell. And indeed, the $\EpJT{10}$
values calculated for \up90Rh with GF and SLGT0 interactions are
smooth (see fig.\ref{fig-EpRh90Investigation}$b$), like in model
spaces without $f_{5/2}$ shell for \up48Cr, see
fig.\ref{fig-EpCr48Investigation}. The SSSV calculated energies behave
in a different way, suggesting the anticipated interaction dependence
of the calculated pairing energy.

There are more cases when different interactions predict similar
pairing energy. For example, the \Pbar{01} pairing energy calculated
along the ground state band in \up48Cr using the modified
surface-delta interaction in the model space $(f_{7/2}p_{1/2})$
\cite{AJSABandTermination} is very similar to the values calculated
using the KB3 interaction in the full $pf$ model space. In addition,
even the decomposition into direct and structural effects
(eq.\refeq{eq-Energy}) gives almost identical values. The MSDI
parameters were fitted to reproduce experimental energies in this
nucleus, thus the model space effect, i.e.\ the absence of $f_{5/2}$,
was probably remedied by this adjustment. The \Pbar{10} pairing energy
calculated with MSDI is smaller than the one obtained using KB3 in the
same model space, shown in fig.\ref{fig-EpCr48Investigation}.

Fitting to the energies, in general, does not imply that all
energy-related properties will be well described. For example, the
\Pbar{01} pairing energy, calculated for \up48Cr using the
surface-delta interaction fitted to the energies in the full $pf$
model space (the parameter values $A_T$=0.2+0.4$T$), is larger than
KB3 values. However, the spin dependence has the same profile.

\begin{figure}[tb]
\centerline{\psfig{figure=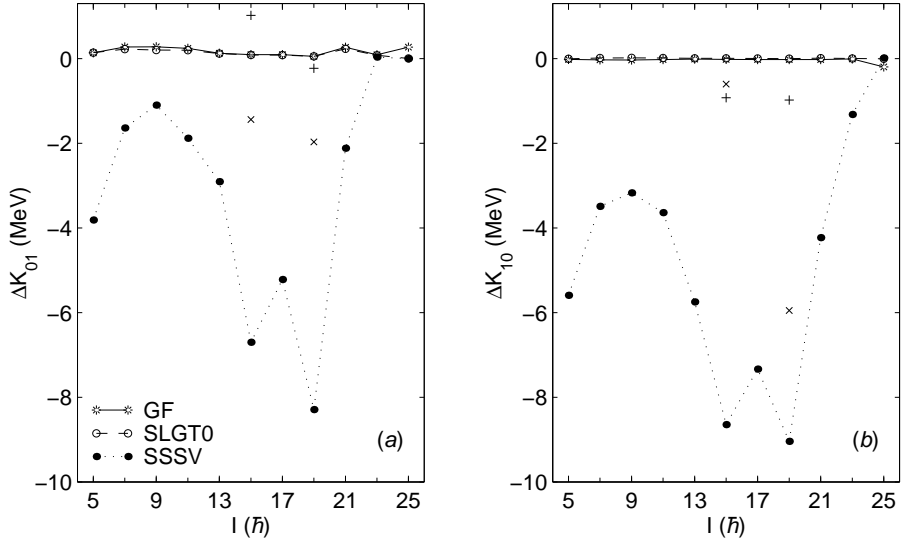,height=\figheight}}
\caption{%
Interaction dependence of the single-particle contribution difference
calculated for \up90Rh nucleus using GF, SLGT0 and SSSV
interactions. In the latter case the single-hole contribution
difference is shown. The left (right) plot shows \Pbar{01} (\Pbar{10})
case. The notation is the same as in
fig.\ref{fig-EpRh90Investigation}.}
\label{fig-KdRh90Investigation}
\end{figure}

The single-particle contribution difference, calculated in \up90Rh
using three different residual interactions, is shown in
fig.\ref{fig-KdRh90Investigation}. As was already seen from the model
space investigation of $\Delta\spt$, the contributions may be
different in these calculations, first of all, because of the
different model spaces.  Thus the $\spt$ values obtained in GF and
SLGT0 interactions, being one similar to another, should be expected
to be different from the the SSSV values. As seen from the figure, the
pairing interactions favour configurations with more holes excited
from the $g_{9/2}$ shell in the SSSV calculation. A similar conclusion
was made for \up48Cr.

\subsubsection{Pairing energy}
\label{ssect-PairingEnergy}

\begin{figure}[tb]
\centerline{\psfig{figure=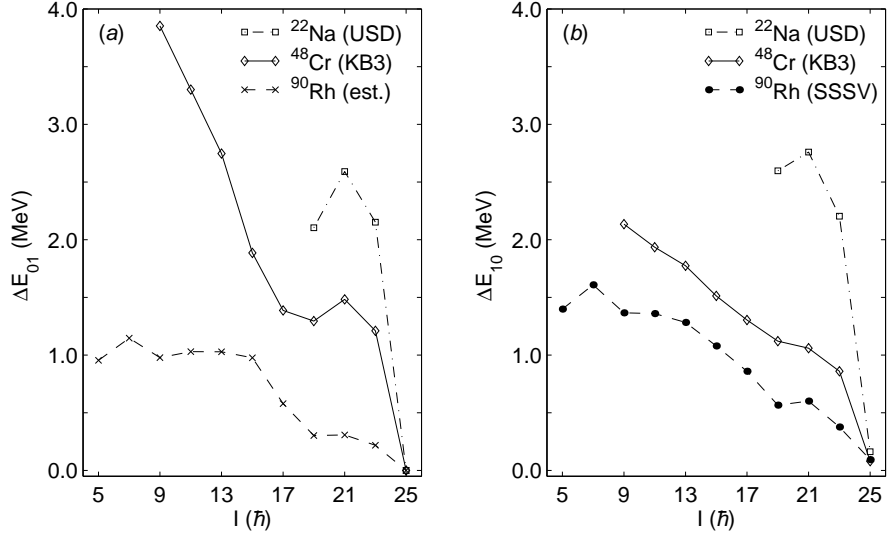,height=\figheight}}
\caption{Spin dependence of the calculated pairing energy 
in the selected bands of \up22Na, \up48Cr and \up90Rh nuclei. For
explanations see text. The \Pbar{01} (\Pbar{10}) case is shown on the
left (right).  }
\label{fig-Epairing}
\end{figure}

The $JT$ pairing energies, calculated for three selected nuclei, are
shown in fig.\ref{fig-Epairing}. The spin values in \up22Na and
\up48Cr calculations are shifted to match the maximum spin value
$\Imax$=25$^+$ of the selected band in \up90Rh by 16 and 9$\hbar,$
respectively. Pairing energies in \up22Na and \up48Cr are shown as
obtained from the calculations using USD and KB3 interactions,
respectively. While for the nucleus \up90Rh, the estimated \Pbar{01}
pairing energy is shown. Estimation tries to account for the
spin-orbit partner of the $g_{9/2}$ shell, namely the $g_{7/2}$ shell,
that is not included in the model space. It is made by assuming that
the pairing energy scales with the model space degeneracy, as
suggested by the model space investigation in
subsection \ref{ssect-ModelSpace}. The estimation is based on the calculated
SSSV pairing energies (since they are obtained in a larger model space
than, say, GF values), thus the scaling parameter is 1.5. The
\Pbar{10} pairing energy in \up90Rh is taken as obtained using the
SSSV interaction, because \Pbar{10} energies do not scale with the
model space. As can be seen from the figure, both the \Pbar{01} and
\Pbar{10} pairing energies decrease as spin increases (on the
average). In addition, the values are similar in magnitude for the
odd-odd nuclei, \up22Na and \up90Rh.  In the \up90Rh calculation using
the SSSV interaction $\EpJT{10}$ is even larger than
$\EpJT{01}$. While in the case of even-even nucleus, \up48Cr, the
influence of $JT$=01 interaction is larger, as was also shown in
ref.\cite{PovesMartinezCr48}. Thus both pairing interactions reduce
the moment of inertia. However, the relative importance of \Pbar{01}
and \Pbar{10} pairings to the moment of inertia in $N$=$Z$ nuclei
depends on whether it is an odd-odd or even-even nucleus; in addition,
it depends on the total isospin of states as well as on the oddity of
state spin. Fig.\ref{fig-Epairing} shows that the spin dependencies of
\Pbar{01} and \Pbar{10} pairing energies are more or less
correlated. Those issues are addressed in a greater detail in
ref.\cite{AJSAGa62} where pairing properties of the rotational bands
in the odd-odd $N$=$Z$ nucleus \up62Ga are discussed.

Due to a spin shift when plotting, it is easier to see that the
irregularities in the spin dependence of the \Pbar{01} pairing
energies are similar in the bands (see
fig.\ref{fig-Epairing}$a$). There is no $JT$=01 coupled pairs in the
aligned, band terminating state, thus there is no pairing energy at
spin $\Imax$. There is a local maximum at spin $\Imax-$4, which
happens to be the backbending point in \up48Cr and \up90Rh (see figs.\
\ref{fig-NaCrEnergies} and \ref{fig-RhEnergies}). This state was also
described as the maximum spin state for seniority $v_\mathrm{max}-$4
(where $v_\mathrm{max}$=2$j$+1), and the discussion in subsection
\ref{ssect-Selection} suggests that the pairing energy should be
smaller in the neighbouring states: at higher spins the total
seniority becomes larger; while at spin $\Imax-$6 seniority is the
same as in state $\Imax-$4 in a single-$j$ picture, however, this
state is less resistant to the shell mixing that breaks $JT$=01 pairs,
since it is not ``maximum spin for a given seniority'' state
\cite{AJSABandTermination}. 
The \Pbar{01} pairing energy increases as spin decreases from
$\Imax-$6 to $\Imax-$10, and continues to increase for \up48Cr towards
the lower spins, but remains rather constant for \up90Rh. It may be an
effect from the $T$=1 pairing blocking in $\Ttot$=0 states of an
odd-odd nucleus.

The spin shift reveals also similarities of the $\EpJT{10}$
spin-behaviour when values are compared starting from the band
terminating state, see fig.\ref{fig-Epairing}$b$. The $\Imax-$4 
state has an irregularity as compared with the trend of lower spin
states. For the spins below $\Imax-$4, the $\EpJT{10}$ is almost a
linear function of spin. In \up48Cr there is a slight increase of
$\EpJT{10}$ (as compared with this linear interpolation) at spin
$\Imax-$12=4$^+.$ This spin value is the lowest state if only
\Pbar{10} pairing interaction constitutes the complete
interaction. Thus it may be a signature, that KB3 contains
overestimated \Pbar{10} force (cf.\ fig.\ref{fig-SingleJEp}), because
a similar irregularity is also seen in the full energy spectrum with a
rotational reference subtracted, while there is no such irregularity
in experimental data (see fig.\ref{fig-NaCrEnergies}). 

\begin{figure}[tb]
\centerline{\psfig{figure=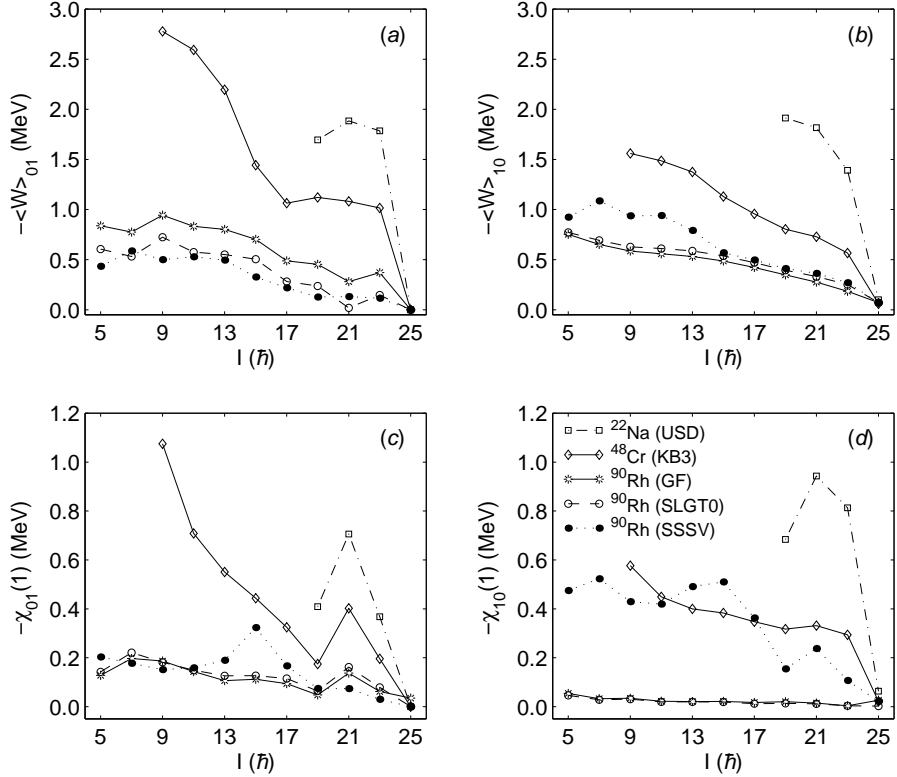,width=12cm}}
\caption{
Parameterization coefficients of the pairing energy, see
eqs.\refeq{eq-Energy} and \refeq{eq-PairingEnergy}. Details are given
in text. The left (right) column shows \Pbar{01} (\Pbar{10}) case.}
\label{fig-Coeffs}
\end{figure}

The parameter values in the decomposition of \Pbar{JT} pairing
energies, see eqs.\refeq{eq-Energy} and \refeq{eq-PairingEnergy}, are
shown in fig.\ref{fig-Coeffs}: plots in the top row show expectation
values of the normalized pairing interaction $\Wav{JT}$, while the
bottom row plots show the total structural effect $\chi_{JT}$(1).
Like in fig.\ref{fig-Epairing}, the spin values for \up22Na and
\up48Cr are shifted. The parameter values for \up22Na and \up48Cr are
calculated using USD and KB3 interactions, respectively. For the
\up90Rh nucleus, parameter values obtained using all three
interactions, GF, SLGT0 as well as SSSV, are given. The values are not
scaled to estimate the effect from the excluded $g_{7/2}$ shell.
The irregularities of $\chi$(1) values calculated for
\up90Rh may be attributed to the two state interaction, discussed in
section \ref{ssect-Wavefunction}. To eliminate the effect
coming from particles in the $p_{1/2}$ shell, the $\Wav{JT}$ values
obtained in the GF as well as SLGT0 calculations are shifted down to
match the SSSV value at $I$=25$^+$ (cf.\
fig.\ref{fig-EpRh90Investigation}).  The $\chi(1)$ values show that
the non-linear term, and thus the reconfiguration effect, is not
negligible and plays an important role in shaping the spin dependence
of the bands. In particular, there is an increase in the nonlinear
contribution at spin ($\Imax-$4)$\hbar$. The difference in the
magnitude of the parameter values in \up22Na, \up48Cr and \up90Rh
cannot be explained by simple mass scaling, $\hbar w.$ There are
additional effects, that may have origin in the dependence on the
residual interaction.

As seen from the fig.\ref{fig-Coeffs}$a$, the expectation value of the
\Pbar{01} interaction is approximately constant for spins ranging from
$\Imax-$2 to $\Imax-$6 (or even $\Imax-$8), suggesting that the
average number of $JT$=01 pairs in this spin range is not changing in
the Hamiltonian without \Pbar{01} pairing interaction. The
fig.\ref{fig-Coeffs}$c$ suggests, that the structural effect is
responsible for the increase of the pairing energy at spin $\Imax-$4.
The figs.\ref{fig-Coeffs}$a$ and $c$ suggest, that the unfavoured
band termination is due to the equal number of $JT$=01 pairs at spins
$I$=$\Imax-$4 and $\Imax-$2 in the calculation without this pairing
interaction, since the structural effect changes by the same amount
when going from $\Imax-$4 to $\Imax-$2, and finally $\Imax$ state. The
$T$=0 pairing helps in making the $\Imax$ state unfavourable: here
the change in the pairing energy is larger when going from $\Imax-$2
to $\Imax$ state as compared to the change between states $\Imax-$4
and $\Imax-$2.

The structural change is also important for the effect coming from the
\Pbar{10} interaction (see plots in the right-column of
fig.\ref{fig-Coeffs}). The expectation value of the pairing
interaction has smoother spin dependence than the pairing energy does
in the odd-odd nuclei (cf.\ fig.\ref{fig-Epairing}). It is interesting
to note, that the difference in $\EpJT{10}$ between the GF (or SLGT0)
and SSSV interactions is coming from the structural effects, while the
average number of $L$=0, $S$=$J$=1, $T$=0 pairs is the same at
$\lambda$=0. 

\section{Summary and conclusions}
\label{sect-Summary}

There are several ambiguities, when discussing the role of the pairing
interaction in the shell model. First problem is the ``no-pairing''
Hamiltonian. Since it is defined as a deviation from the full
Hamiltonian (see eq.\refeq{eq-NoPairingHam}), the correctness of
defining the magnitude of this deviation (i.e.\ the pairing strength)
is crucial to have a correct Hamiltonian without pairing. Having
selected (or fixed from other considerations) the strength of the
pairing interactions, a question may be asked whether a {\em full\/}
force is present in a chosen residual interaction. This question was
also discussed in ref.\cite{DufourZuker}, where it was argued that the
multipole part of realistic interaction, that is responsible for
particle correlations, including pairing interactions, is almost the
same. Thus only small deviations from the pairing interaction content,
estimated in this reference, are expected. However, the model space
dependence of the pairing properties may yield unwanted effects.

Another problem is the form of the pairing interaction. The normalized
$JT$=01 pairing interaction was shown \cite{DufourZuker} to remedy the
divergence in the pairing plus quadrupole model, in the case when many
shells are included in the model space. However, this normalized form
still does not solve the divergence problem completely. Thus the form
of the force is not final yet, and details of the derived role of
pairing interactions may change.

The third problem is the interpretation of the role of a pairing
interaction in comparison with that derived from other models. 
The difference in treatment of pairing correlations
in the shell model and, say, Hartree-Fock-Bogolyubov approach puts a
question whether comparison is meaningful.

The fourth problem is the relation between the pairing properties
obtained in the shell model and observables.

Having the above ambiguities in mind, lets summarize the presented
investigation of the role of pairing interactions.  Three selected
bands in \up22Na, \up48Cr and \up90Rh, having the total isospin
$\Ttot$=0, were discussed. Their choice, based on the behaviour of the
$JT$=01 pairing energy in a single-$j$ shell model, was explained.
Since there is no experimental data on \up90Rh available, three
interactions were used to calculate its properties, and a comparison
of their results was presented. All three interactions predict that
the ground state is $I$=0, $\Ttot$=1, and that the excitation energy
to the lowest $\Ttot$=0 states is approximately 1.3 MeV.  In addition,
the $K^\pi$=5$^+$ ($\Ttot$=0) band, which was of interest in the
present investigation, backbends at spin $I$=19$^+.$ In the SSSV
calculation, the backbending is related to a bandcrossing. The SSSV
calculation should describe quadrupole properties more reliable, thus
odd-$I$ members of the $K$=5$^+$ band are expected to have a prolate
deformation. The comparison of the energies calculated in \up90Rh
using the GF (or SLGT0) and SSSV interactions raises a question,
whether the model space used to describe
\up48Cr yrast band is large enough to discuss also the high spin 
states. This question is supported by the observation that the
deviation between calculated and observed energies is increasing after
$I$=10 towards the aligned state at 16$^+$ (see
fig.\ref{fig-NaCrEnergies}). In addition, the proposed shell model
explanation for the backbending as a result of interplay between the
monopole and quadrupole forces \cite{QuasiSU3} does not exclude the
picture of the unpaired band-crossing (ground state band crosses with
the $K^\pi$=8$^+$ band).

A perturbative approach to investigate the role of schematic
normalized $L$=0 pairing interactions \cite{DufourZuker} was
presented. Both $T$=0 and $T$=1 pairing interactions reduce the moment
of inertia (that is in agreement with \cite{PovesMartinezCr48}),
however their relative importance have interesting dependence on the
spin oddity and total isospin of the states (see \cite{AJSAGa62}). It
was shown that the pairing force brings energy in two ways: (i)
directly, due to an extra binding of certain pairs of nucleons, and
(ii) indirectly, due to an increased likelihood to form those pairs. From
the calculations it follows, that the indirect effect is as important
as the direct effect.

Problems occurring during investigation of the role of pairing interactions
were discussed. In particular, the case of two close-lying states was
described. It was shown that the pairing interaction do not alter
quadrupole properties, if the energy distance between the states is
large enough.

The dependence of the pairing energy (the difference between energies
of yrast states calculated from the Hamiltonian without pairing
interaction and the full Hamiltonian) on the model space was
investigated taking \up48Cr nucleus as an example. It was shown that
the spin dependence of the \Pbar{01} pairing energy derived in the
full model space may be estimated from the calculation in a smaller
model space by taking into account the increased degeneracy of the
model space. This observation was used to estimate the $JT$=01 pairing
energy in \up90Rh as would be obtained if the $g_{7/2}$ shell would be
included in the model space. It was also shown that the \Pbar{10}
pairing energy does not scale with the model space degeneracy. The
pairing energy dependence on the choice of the residual interaction
was also discussed.  During the investigations of those two
dependencies, it was shown that the pairing interactions in most cases
facilitate the shell mixing (a similar picture occurs in the BCS
model).

The pairing energies calculated in the three selected bands were
compared. It was shown, that if the spin values are shifted so that
the aligned state having spin $\Imax$=($p$+1)$^2$ coincides, the spin
dependencies of pairing energies in the three selected bands have
similar profiles. The decomposition of the pairing energy into direct
and indirect parts, available due to a perturbative way of
investigation, was also presented. It was shown that the similarities
in the spin dependence of the coefficients remain.

Despite being rather detailed, this investigation raised questions
which could not be answered. For example, the energy spectrum of
\up90Rh remains an open question. The three used interactions predict
different not only excitation energies, but also the energy distance
between the yrast and yrare states. The origin of the backbend of the
$K^\pi$=5$^+$ band in this nucleus is also unclear: whether it is
similar to the one discussed in ref.\cite{QuasiSU3} or it is ``usual''
backbending originating from a band-crossing.  The investigation of
the pairing energy dependence on the model space suggests that the
calculation performed in the model space containing also $g_{7/2}$
shell may give somewhat different results than discussed here. 
The experimental data would provide a stringent test for the choice of
the residual interaction.

In addition, a systematic study of the pairing energy dependence on
the total isospin and the oddity of state spin is needed to add an
understanding what is the origin of blocking effects.

\begin{ack}
The author thanks E.\ Caurier for access to the shell model code
\cite{Antoine}, and S.\ {\AA}berg for useful discussions as well as
comments on the manuscript. The financial support from The Swedish
Institute (``The Visby Programme'') is also acknowledged.
\end{ack}


\end{document}